\documentclass[aps,prb,reprint,amsmath,amssymb,floatfix]{revtex4-2}

\usepackage{graphicx,color,hyperref,overpic}
\usepackage[ignoreunlbld,norefs,nocites]{refcheck}

\usepackage[outdir=./]{epstopdf}

\begin{document}

\title{Structures of bulk hexagonal post-transition-metal
  chalcogenides from dispersion-corrected density-functional theory}

\author{S.\ J.\ Magorrian}

\affiliation{National Graphene Institute, University of Manchester,
  Booth Street East, Manchester M13 9PL, United Kingdom}

\author{V.\ Z\'{o}lyomi}

\affiliation{Hartree Centre, STFC Daresbury Laboratory, Daresbury WA4
  4AD, United Kingdom}

\author{N.\ D.\ Drummond}

\affiliation{Department of Physics, Lancaster University, Lancaster
  LA1 4YB, United Kingdom}

\date{\today}

\begin{abstract}
We use dispersion-corrected density-functional theory to determine the
relative energies of competing polytypes of bulk layered hexagonal
post-transition-metal chalcogenides, to search for the most stable
structures of these potentially technologically important semiconductors.
We show that there is some degree of consensus among
dispersion-corrected exchange-correlation functionals regarding the
energetic orderings of polytypes, but we find that for each material
there are multiple stacking orders with relative energies of less than
1 meV per monolayer unit cell, implying that stacking faults are
expected to be abundant in all post-transition-metal chalcogenides.
By fitting a simple model to all our energy data, we predict that the
most stable hexagonal structure has P$6_3$/mmc space group in each
case, but that the stacking order differs between GaS, GaSe, GaTe,
and InS on the one hand and InSe and InTe on the other.
At zero pressure, the relative energies obtained with different
functionals disagree by around 1--5 meV per monolayer unit cell, which
is not sufficient to identify the most stable structure unambiguously;
however, multi-GPa pressures reduce the number of competing phases
significantly.
At higher pressures, an AB$'$-stacked structure of the most stable
monolayer polytype is found to be the most stable bulk structure; this
structure has not been reported in experiments thus far.
\end{abstract}

\maketitle


\section{Introduction}

\begin{figure}[!htbp]
\centering
\includegraphics[clip,width=0.45\textwidth]{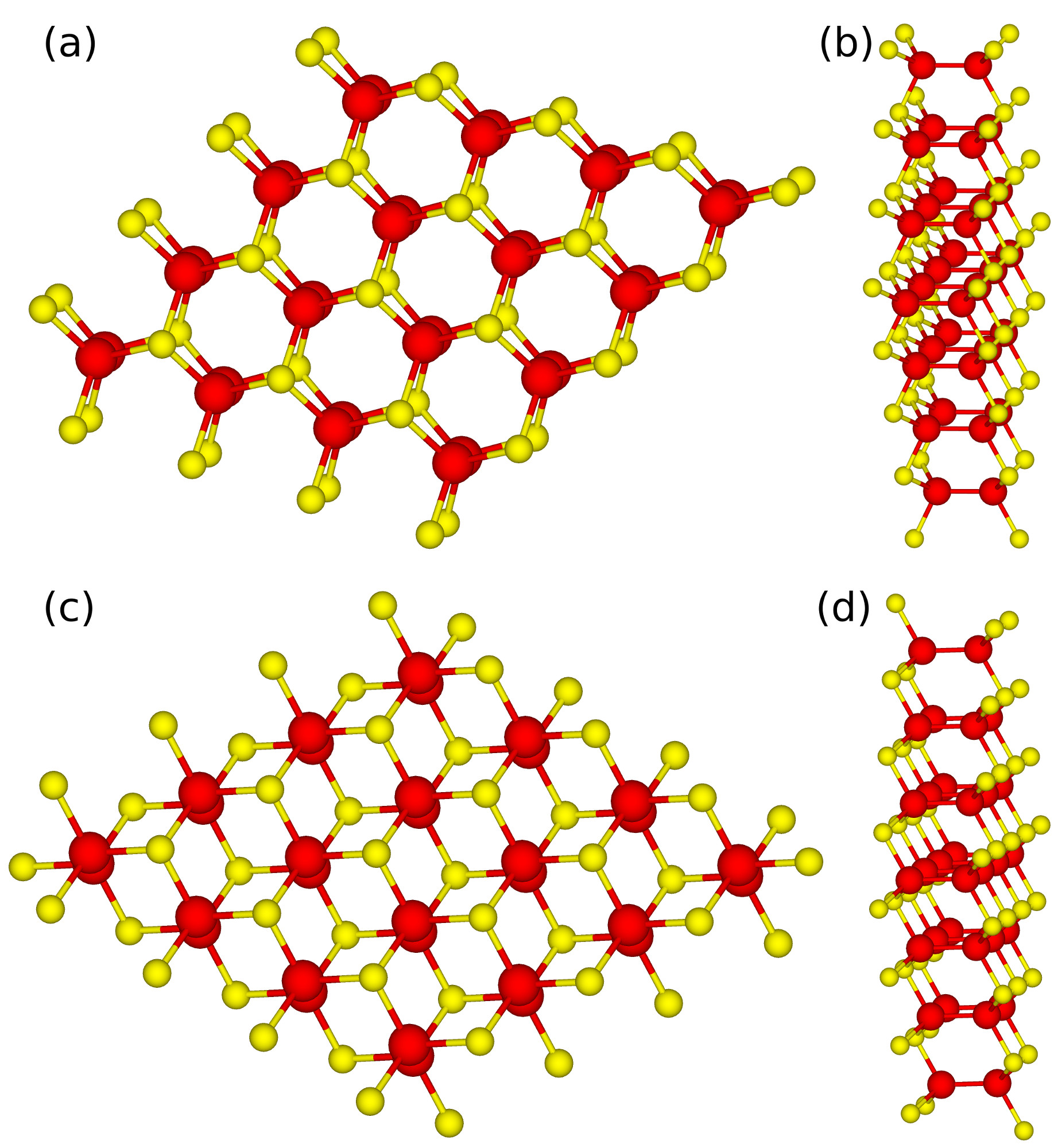}
\caption{(a) Top and (b) side views of the $\alpha_{\rm M}$ polytype
  of monolayer GaS, and (c) top and (d) side views of the $\beta_{\rm
    M}$ polytype \cite{Zolyomi_2013,Zolyomi_2014}.
Gallium and sulfur atoms are shown in red and yellow, respectively.
\label{fig:mono_GaS_structure}}
\end{figure}

The hexagonal post-transition-metal chalcogenides (PTMCs) GaS, GaSe,
GaTe, InS, InSe, and InTe are layered materials with hexagonal Bravais
lattices \cite{Schubert_1953,Hahn_1955,Sugaike_1957}.
Due to the possibility of isolating mono- and few-layer films, in
their ultrathin form they have received considerable attention in
recent years as a new class of two-dimensional (2D) semiconductor
\cite{Late_2012,Hu_2012,Zolyomi_2013,Zolyomi_2014,Tamalampudi_2014,
  Liu_2014,Cao_2015,Mudd_2015,Sucharitakul_2015,Bandurin_2016,Aziza_2017,
  Hung_2017,Terry_2018,Hamer_2019}.
The two dynamically stable structures of PTMC monolayers are shown in
Fig.\ \ref{fig:mono_GaS_structure}, and are based on the honeycomb
motif \cite{Zolyomi_2013,Zolyomi_2014}.
Bulk PTMCs have direct band gaps of $\sim 1.3$--$2.5$ eV
\cite{Voitchovsky_1974,Camassel_1978, Alekperov_1991, Ho_2006}, light
out-of-plane effective masses
\cite{Tredgold_1969,Ottaviani_1974,Schluter_1976, Kuroda_1980,
  KressRogers_1982, Gomes_1993}, and strongly nonlinear optical
properties such as second harmonic generation, optical gain, and
up-/down-conversion \cite{Fernelius_1994, Cingolani_1981, Segura_1997,
  Singh_1998, Allakhverdiev_2009}.
InSe exhibits high in-plane electron mobility \cite{Segura_1984},
which persists in the thin-film limit, and has enabled the observation
of the quantum Hall effect \cite{Bandurin_2016} and the demonstration
of PTMCs as candidate ultrathin transistors
\cite{Late_2012,Sucharitakul_2015}.
InSe has also shown potential for applications in photovoltaics
\cite{Segura_1979, Segura_1983} and electron-beam-based data-storage
\cite{Gibson_2005}.

Thin films of PTMCs exhibit high-sensitivity broadband photoresponse
\cite{Hu_2012, Tamalampudi_2014, Liu_2014, Mudd_2015}.
They also show a substantial increase in the band gap, from $1.3$ eV
in bulk InSe to $\sim 2.8$ eV in monolayer InSe \cite{Bandurin_2016},
and from $2$ eV in bulk GaSe to $\sim 3.5$ eV in monolayer GaSe
\cite{Aziza_2017, Terry_2018}.
An offset in the location of the valence-band maximum has been shown
to develop in the thinnest films \cite{Aziza_2018, Hamer_2019},
yielding a slightly indirect band gap, unlike the bulk.
Combined with the high density of states at the band edge, this is
expected to lead to strongly correlated phenomena in p-doped monolayer
PTMCs \cite{Zolyomi_2013, Zolyomi_2014, Cao_2015} as well as
interesting thermoelectric properties \cite{Hung_2017}.

The high tunability of the physical properties of PTMC films stems
from the strong electronic coupling between states localized on
neighboring layers \cite{Magorrian_2016}.
For this reason, PTMCs are likely to be highly sensitive to changes
brought about by variations in stacking order.
The influence of stacking and interlayer interactions has already been
shown to be important in, for example, the metallic transition-metal
dichalcogenides \cite{Ritschel_2018, Lee_2019, Stahl_2020}, which
feature multiple stacking orders very close in energy.
The local stacking order will vary continuously in the
moir{\'e} superlattices formed when monolayers are stacked with a
relative rotation or lattice-constant mismatch.
In twisted bilayers of 2D materials with small misalignments and/or
lattice-constant mismatches, the constituent monolayers can adjust to
maximize the size of regions of energetically favorable stacking
\cite{Enaldiev_2020, Weston_2020}.
Compared to graphene and transition-metal dichalcogenides, PTMCs have
low Young's moduli and are highly flexible \cite{Chitara_2018,
  Zhao_2019, Demirci_2017}, so in-plane relaxation can be expected to
occur more readily, starting from larger twist angles, and featuring
stronger nonuniform strain fields, than in twisted
transition-metal-dichalcogenide bilayers.
To describe such reconstruction in moir{\'e} superlattices of PTMCs it
is essential first to attain a proper understanding of the energetics
of the various PTMC polytypes and structures, and the factors
contributing to their formation.

In this work, we use a range of dispersion-corrected
density-functional-theory (DFT) methods to investigate systematically
the energies and stabilities of the competing polytypes of bulk
layered hexagonal PTMCs.
We provide an expression for the energy per monolayer unit cell of an
arbitrary bulk hexagonal PTMC polytype.
We find that each PTMC generally admits a few polytypes that are
energetically very similar, implying that crystal-growth conditions
are likely to be important.
Motivated by the observation of electronic and structural changes in
PTMCs under pressure
\cite{dAmour_1982,Ulrich_1996,Errandonea_1999,Errandonea_2005}, we
also investigate the pressure-dependence of the relative stability of
competing polytypes.

The post-transition-metal (PTM) atoms that we consider are indium and
gallium, while the chalcogen atoms that we consider are sulfur,
selenium, and tellurium.
The PTM atoms are strongly bonded in vertical dimers lying on a
hexagonal sublattice.
Each PTM atom is strongly bonded to three chalcogen atoms lying on a
different hexagonal sublattice to the PTM dimers.
There are two different single-layer polytypes, as shown in
Fig.\ \ref{fig:mono_GaS_structure}: the chalcogen atoms may all lie on
the same sublattice, or the top and bottom chalcogen atoms may lie on
different sublattices \cite{Zolyomi_2013,Zolyomi_2014}.
The former structure, referred to as the $\alpha_{\rm M}$ monolayer
polytype, is slightly more stable and has vertical mirror symmetry
$\sigma_h$ about the center of the layer, although it lacks inversion
symmetry (D$_{3h}$ point group).
The latter structure, referred to as the $\beta_{\rm M}$ monolayer
polytype, does not have vertical mirror symmetry, but it does have
inversion symmetry (D$_{3d}$ point group).
In bulk hexagonal PTMCs there are further possibilities for polytypism
due to the different ways in which the layers can be stacked.
Our reference structure is the simplest possible bulk structure, which
consists of AA-stacked $\alpha_{\rm M}$-PTMC monolayers, with a
four-atom primitive unit cell.

A range of polytypes and stacking orders have been reported for the
bulk structures of the PTMCs obtained in experiments
\cite{Gouskov_1982}.
The $\beta$ \cite{Kuhn_1976} and $\varepsilon$ \cite{Kuhn1975b} 2H
polytypes both have $\sigma_h$ reflection symmetry, with the former
also having an inversion center.
Meanwhile, the $\gamma$ 3R polytype \cite{Rigoult_1980} has a
single-layer primitive unit cell, and has neither inversion nor
$\sigma_h$ reflection symmetry.
A polytype known as $\delta$, consisting of a four-layer unit cell with
two interfaces between successive layers stacked as in the $\beta$
polytype and the other two interfaces stacked as in the $\gamma$
polytype, has also been reported for GaSe \cite{Kuhn_1975}.
Note that the $\beta_{\rm M}$ monolayer polytype should not be
confused with the $\beta$ polytype of bulk PTMCs: the former refers to
the inversion-symmetric monolayer shown in
Figs.\ \ref{fig:mono_GaS_structure}(c) and
\ref{fig:mono_GaS_structure}(d), the latter to the bulk crystal in
which non-inversion-symmetric monolayers [the $\alpha_{\rm M}$
  monolayer polytype shown in Figs.\ \ref{fig:mono_GaS_structure}(a)
  and \ref{fig:mono_GaS_structure}(b)] stack into an AB-type bulk
crystal that now exhibits inversion symmetry.
To avoid confusion, in Sec.\ \ref{sec:charcode} we adopt a
notation for PTMC stacking that enables unambiguous characterization
of all PTMC crystals irrespective of the monolayer polytypes or
stacking order of successive layers.

The rest of this paper is structured as follows.
Our approach for enumerating physically relevant PTMC structures is
described in Sec.\ \ref{sec:charcode}.
We present a fitting function to describe the energetics of PTMC
polytypes in Sec.\ \ref{sec:fit}.
We compare the DFT energies of PTMC polytypes obtained with different
exchange--correlation functionals in Sec.\ \ref{sec:comparison_fnals}.
Our analysis of the most stable polytypes, including the effects of
pressure, is presented in Sec.\ \ref{sec:stability}.
We examine the relationship between the electronic band gap and the
energetic stability of polytypes in Sec.\ \ref{sec:band_structure}.
Finally, we draw our conclusions in Sec.\ \ref{sec:conclusions}.
Our DFT simulation parameters can be found in Appendix
\ref{app:methodology}.

\section{Characterization of structures}\label{sec:charcode}

The bulk hexagonal PTMC geometries we have examined are as follows.
(i) We assume that each sublayer of chalcogen atoms and each sublayer
of vertical PTM dimers lie at the A, B, or C hexagonal sublattice
sites, because energy minima are overwhelmingly likely to occur at
these high-symmetry configurations.
(ii) We assume that each chalcogen sublayer lies on a different
hexagonal sublattice to the PTM sublayer; our DFT calculations for
InSe confirm that the energy is around 2 eV per monolayer unit cell
higher each time the chalcogen atoms are on the same sublattice as the
PTM dimers.
In general a two-layer structure is a 2H polytype in Ramsdell notation
\cite{Ramsdell_1947}, and a three-layer structure is a 3H polytype.
However, there are exceptions; e.g., the $\gamma$ structure is a 3R
polytype with a rhombohedral primitive Bravais lattice.
Nevertheless, for consistency and ease of automation, we have used a
hexagonal unit cell in all our calculations.

\begin{figure}
\centering
\includegraphics[clip,width=0.48\textwidth]{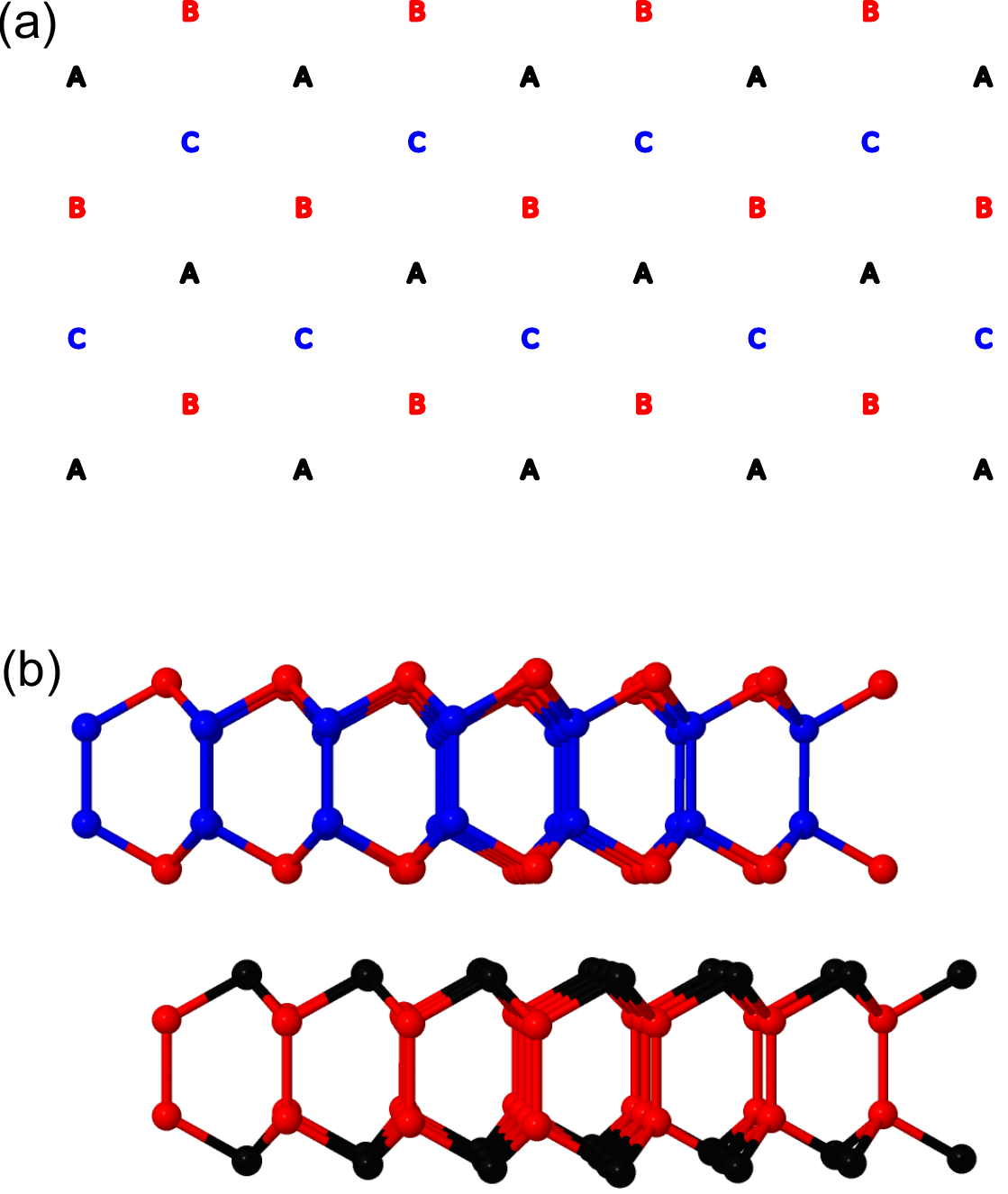}
\caption{(a) 2D hexagonal sublattice labels A, B,
  and C for each sublayer, used to construct structure label
  strings for bulk PTMCs.
(b) Color-coded structure of the
  ``\texttt{a\textcolor{red}{B}a\textcolor{red}{b}\textcolor{blue}{C}\textcolor{red}{b}}''-stacked
  $\varepsilon$-GaSe as an example.
The colors of the atoms in panel (b) correspond to the colors of the
sublattice sites in panel (a).
\label{fig:label_diag}}
\end{figure}

We have performed DFT calculations for all such two-layer and
three-layer bulk structures.
We refer to each of these configurations by a character string
summarizing the 2D hexagonal sublattice sites for each sublayer of
atoms in the unit cell.
Upper-case letters (\texttt{A}, \texttt{B}, and \texttt{C}) are used
for PTM-dimer sublayers; lower-case letters (\texttt{a}, \texttt{b},
and \texttt{c}) are used for chalcogen sublayers.
The 2D hexagonal sublattice sites for a single sublayer are shown in
Fig.\ \ref{fig:label_diag}.
For example, the string ``\texttt{aBabCa}'' describes a two-layer bulk
structure in which the PTM dimers lie on the B and C sublattices,
while the chalcogen atoms in the first layer are all at the A
sublattice sites and the chalcogen atoms in the second layer are at
the B and A sublattice sites.
In this notation, the $\varepsilon$ polytype \cite{Kuhn1975b} is
\texttt{aBabCb}, the $\beta$ polytype \cite{Kuhn_1976} is
\texttt{aBabAb}, the $\gamma$ polytype \cite{Rigoult_1980} is
\texttt{aBabCbcAc}, and the $\delta$ polytype \cite{Kuhn_1975} is
\texttt{aBabAbaCacAc}.

PTMC structures are energetically invariant if we perform any rigid
operations (translations, rotations, or reflections).
In-plane translations from one sublattice to another correspond to
even permutations of the sublattice labels A, B, and C; thus, e.g.,
``\texttt{aBabCa}'' is energetically equivalent to
``\texttt{bCbcAb}.''
In-plane point rotations through $60^\circ$ or reflections in vertical
planes, together with translations, correspond to odd permutations of
the sublattice labels; thus ``\texttt{aBabCa}'' is equivalent to
``\texttt{aCacBa}.''
PTMC structures are also equivalent under vertical displacements,
which correspond to rotating the structure strings through three
characters; thus ``\texttt{aBabCa}'' is equivalent to
``\texttt{bCaaBa}.''
Finally, structures are energetically invariant under reflections in
horizontal planes; thus ``\texttt{aCacBa}'' is equivalent to
``\texttt{aBcaCa}.''

A program was written to loop over all valid structure strings (i.e.,
strings in which each chalcogen atom is at a different sublattice site
to the neighboring PTM dimer) for multilayer bulk structures.
Energetically equivalent structure strings were eliminated and DFT
input files for the remaining structures were generated.
We find that there are $2$ inequivalent one-layer structures (these
being the $\alpha_{\rm M}$ and $\beta_{\rm M}$ polytypes with AA
stacking), $12$ inequivalent two-layer structures (two of these being
supercells of the one-layer structures), $62$ inequivalent three-layer
structures, $494$ inequivalent four-layer structures, $4292$
inequivalent five-layer structures, and $42158$ inequivalent six-layer
structures.
The atomic positions and lattice vectors were relaxed within DFT at
zero external pressure, subject to the constraint of the initial
symmetry.
The imposition of symmetry constrains the unit cell to be hexagonal
and constrains the atoms to 2D hexagonal sites, but it allows the
sublayers to relax in the out-of-plane direction and it also allows
the $a$ and $c$ hexagonal lattice parameters to relax.

\section{Fit to the bulk PTMC energies \label{sec:fit}}

To represent the energy of each structure $S$ we fit
\begin{eqnarray} E(S) & = & E_{\rm c}+\frac{1}{N_{\rm l}(S)} \left[ n_{\rm nc}(S)
E_{\rm nc} + n_{\rm np}(S) E_{\rm np} \right. \nonumber \\ &
& \hspace{5.5em} \left. {} + n_{\rm ab}(S) E_{\rm ab} + n_{\rm snn}(S)
E_{\rm snn} \right]~~~ \label{eq:Etot_fit} \end{eqnarray} to the
energy $E$ per monolayer unit cell, where $N_{\rm l}(S)$ is the number
of PTMC monolayers in structure $S$, $n_{\rm nc}(S)$ is the number of
places in the unit cell in which neighboring chalcogen atoms are on
different hexagonal sublattice sites, $n_{\rm np}(S)$ is the number of
places in the unit cell in which PTM dimers in neighboring layers are
on different hexagonal sites, $n_{\rm ab}(S)$ is the number of
$\beta_{\rm M}$-polytype layers in the unit cell, and $n_{\rm snn}(S)$
is the number of places in which the next-nearest chalcogen atom is on
the same hexagonal site as a PTM dimer.
For our \texttt{aBa} reference structure (AA-stacked $\alpha_{\rm
  M}$-PTMC), $n_{\rm nc}(S) = n_{\rm np}(S) = n_{\rm ab}(S) = n_{\rm
  snn}(S)=0$.
Hence the fitting parameter $E_{\rm c}$ describes the total energy per
monolayer unit cell of the \texttt{aBa} structure.
$E_{\rm nc}$ is the energy associated with neighboring chalcogen atoms
lying on different hexagonal sublattice sites rather than the same
sublattice site.
$E_{\rm np}$ is the energy associated with PTM dimers in neighboring
layers not lying on the same sublattice.
$E_{\rm ab}$ is the energy of the $\beta_{\rm M}$ polytype of a
single layer relative to the energy of the $\alpha_{\rm M}$ polytype.
Finally, $E_{\rm snn}$ is the energy associated with
second-nearest-neighbor chalcogen atoms lying on the same hexagonal
site rather than different sublattice sites.
The energy of structure $S$ relative to the \texttt{aBa} structure is
$E_{\rm rel}(S)=E(S)-E_{\rm c}$.

\begin{figure*}[!htbp]
\centering
\includegraphics[clip,scale=1]{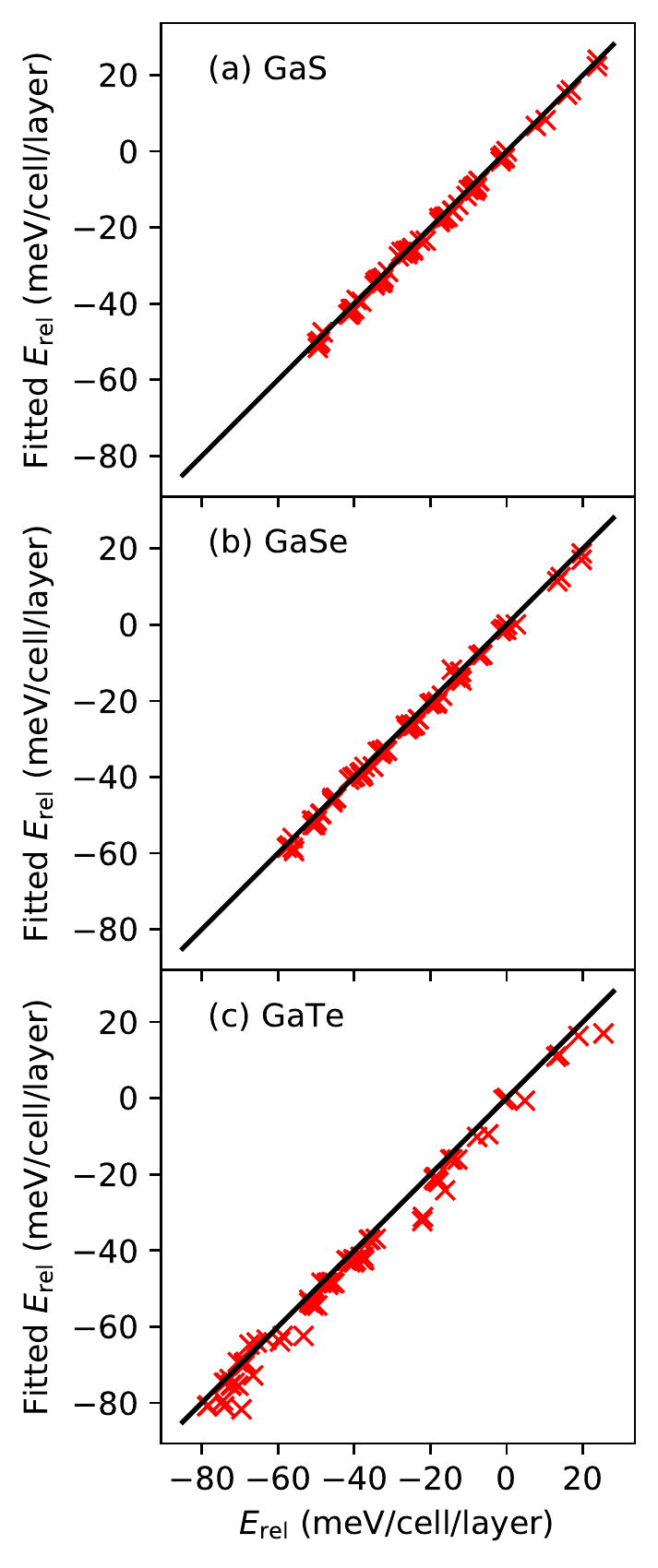} \hspace{2em}
\includegraphics[clip,scale=1]{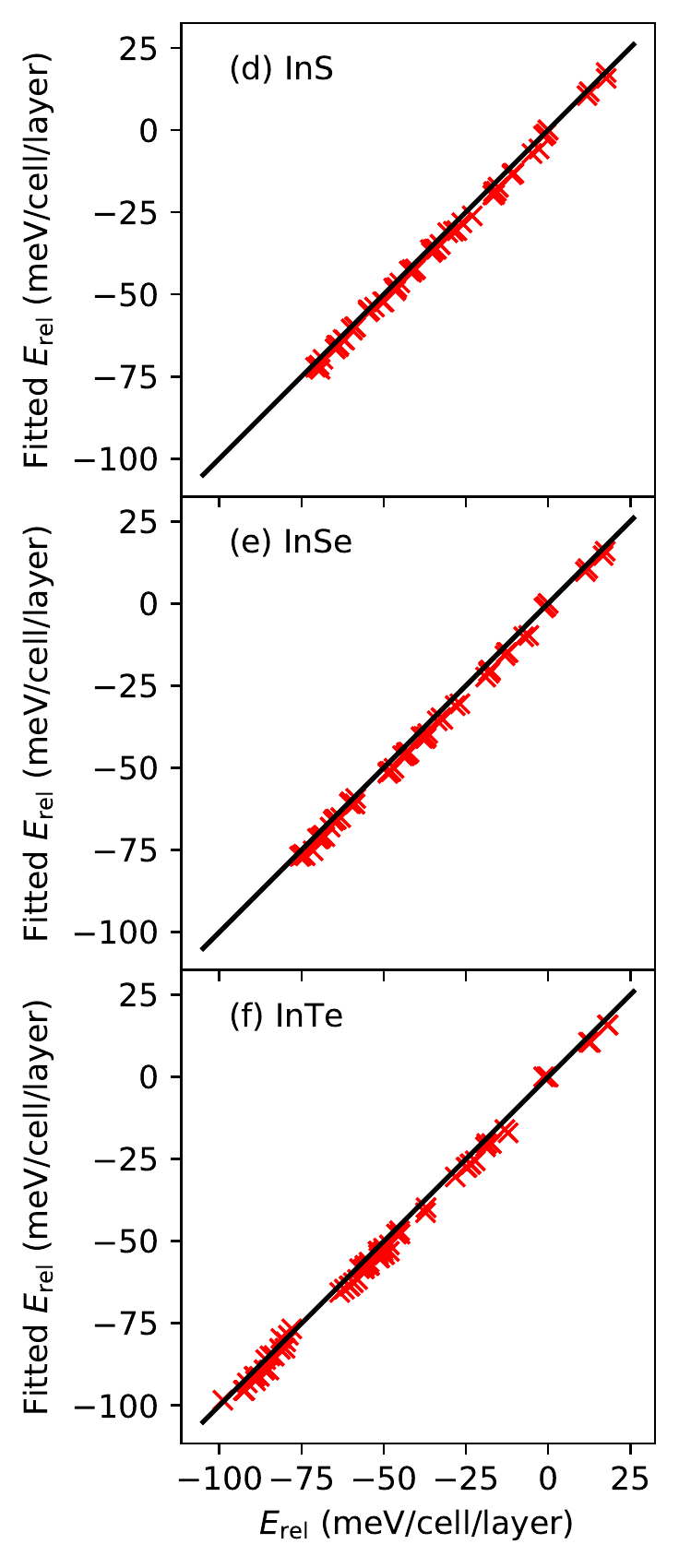}
\caption{Scatter plots showing the fit of Eq.\ (\ref{eq:Etot_fit}) to
  the DFT-PBE-MBD* energy data for (a)--(c) gallium chalcogenides and
  (d)--(f) indium chalcogenides.
$E_{\rm rel}$ is the energy relative to the AA-stacked $\alpha_{\rm
    M}$-PTMC structure [\texttt{aBaaBaaBa}
    ($={}$\texttt{aBaaBa}${}={}$\texttt{aBa})].
  \label{fig:Efit_v_E}}
\end{figure*}

The quality of the resulting fits is illustrated for the DFT-PBE-MBD*
data in Fig.\ \ref{fig:Efit_v_E}.
The fitted parameters and the root-mean-square (RMS) error in the fit
per degree of freedom are reported in Table
\ref{table:MBDstar_fitting_parameters}.
The two- and three-layer structures are all distinct, with the
sole exception of the \texttt{aBa} and \texttt{aBc} structures.
These were independently relaxed for the two- and three-layer
cases, and both the two- and three-layer versions were included
in the fit.
The DFT-PBE-MBD* energy difference between the equivalent two- and
three-layer \texttt{aBa} and \texttt{aBc} structures is around
$1$--$2$ meV per monolayer unit cell, suggesting that the data suffer
from a random error of this order of magnitude due to the finite ${\bf
  k}$-point sampling grids and uncertainties in the relaxed
geometries.
Thus the RMS errors shown in Table
\ref{table:MBDstar_fitting_parameters} are primarily due to noise in
the data rather than any shortcoming in the fitting function of
Eq.\ (\ref{eq:Etot_fit}).
The fit to the GaTe and InTe energy data is clearly significantly
poorer than for the other PTMCs.

\begin{table}[!htbp]
\centering
\caption{Parameters in the fit of Eq.\ (\ref{eq:Etot_fit}) to our two-
  and three-layer DFT-PBE-MBD* PTMC energy data, together with the RMS
  error per degree of freedom.
The parameters and the RMS error are in units of meV per monolayer
unit cell.
Parameter $E_{\rm c}$ in Eq.\ (\ref{eq:Etot_fit}) (the total energy of
the \texttt{aBa} structure) contains an arbitrary,
pseudopotential-dependent offset, and is therefore not reported here.
\label{table:MBDstar_fitting_parameters}}
\begin{tabular}{lr@{.}lr@{.}lr@{.}lr@{.}lr@{.}l}
\hline \hline

PTMC & \multicolumn{2}{c}{$E_{\rm nc}$} & \multicolumn{2}{c}{$E_{\rm
    np}$} & \multicolumn{2}{c}{$E_{\rm ab}$} &
\multicolumn{2}{c}{$E_{\rm snn}$} & \multicolumn{2}{c}{RMS error} \\

\hline

GaS & $-47$&$529$ & $-1$&$709$ & $24$&$028$ & $-1$&$202$ &
~~~~~$1$&$11$ \\

GaSe & $-56$&$010$ & $-1$&$692$ & $18$&$742$ & $-0$&$857$ & $1$&$02$
\\

GaTe & $-79$&$411$ & $-0$&$662$ & $17$&$002$ & $-0$&$804$ & $2$&$04$
\\

InS & $-69$&$786$ & $-1$&$793$ & $17$&$499$ & $-0$&$621$ & $0$&$842$
\\

InSe & $-76$&$943$ & $-1$&$347$ & $15$&$917$ & $1$&$513$ & $1$&$09$ \\

InTe & $-98$&$382$ & $0$&$041$ & $15$&$794$ & $2$&$931$ & $3$&$11$ \\

\hline \hline
\end{tabular}
\end{table}

\section{Comparison of DFT functionals \label{sec:comparison_fnals}}

We have computed the DFT energies within the local density
approximation (LDA) and the Perdew-Burke-Ernzerhof (PBE) variant of
the generalized gradient approximation \cite{Perdew_1996}.
We compare a representative set of semiempirical dispersion-correction
schemes: Grimme 2006 (G06) \cite{Grimme_2006}; Ortmann, Bechstedt, and
Schmidt (OBS) \cite{Ortmann_2006}; and the many-body dispersion (MBD*)
method \cite{Tkatchenko_2012,Ambrosetti_2014}.
We have also investigated the optB86b (MK) and optB88 (BO) nonlocal
van der Waals density functionals \cite{Klime_2009}.
DFT simulation parameters such as the plane-wave cutoff energy are
summarized in Appendix \ref{app:methodology}.

\begin{figure*}[!htbp]
\centering
\includegraphics[clip,scale=0.6]{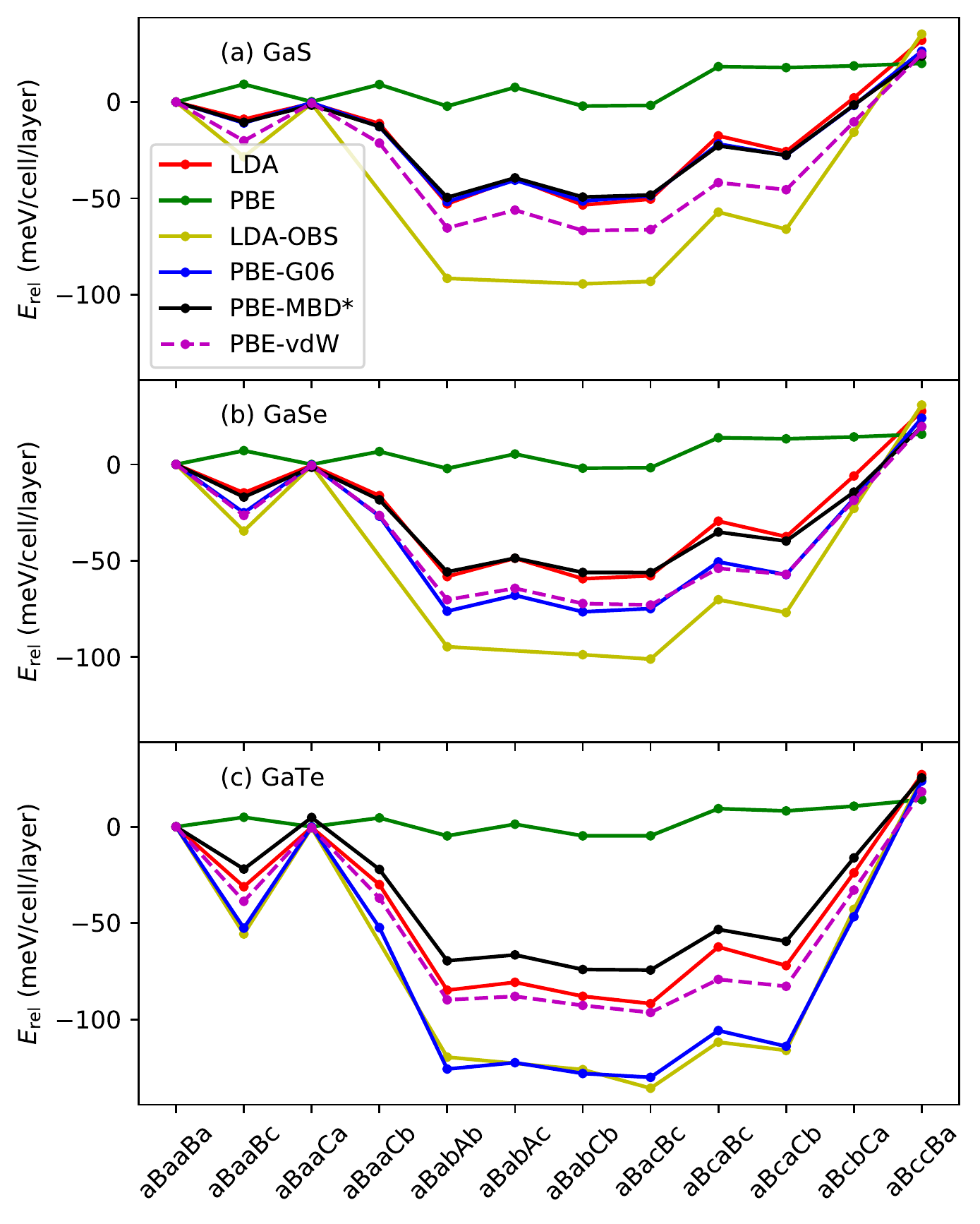}
\includegraphics[clip,scale=0.6]{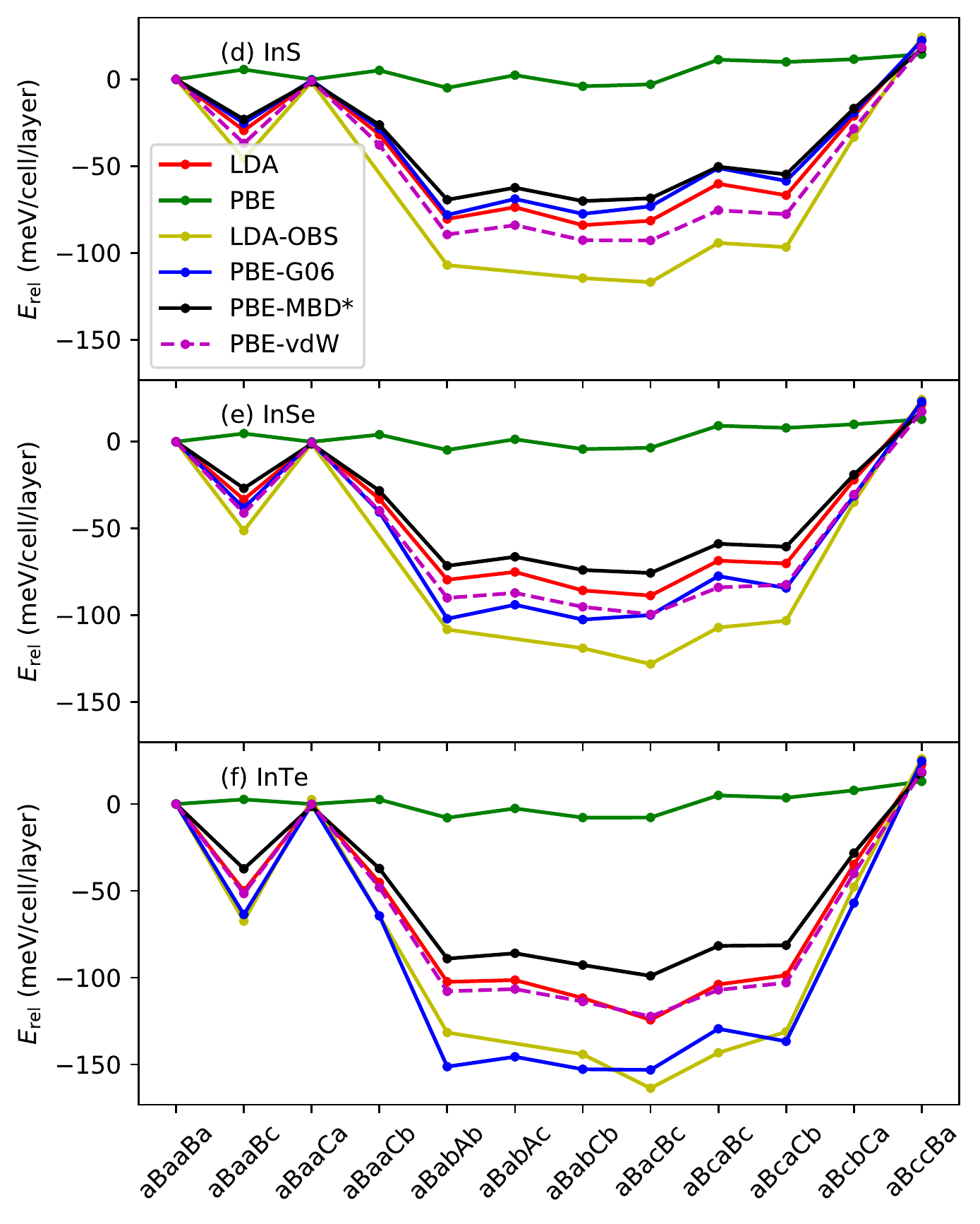}
\caption{DFT energy $E_{\rm rel}$ of two-layer structures relative to
  the AA-stacked $\alpha_{\rm M}$ polytype [\texttt{aBaaBa}
    ($={}$\texttt{aBa})], for (a) GaS, (b) GaSe, (c) GaTe, (d) InS,
  (e) InSe, and (f) InTe.
Different exchange-correlation functionals and dispersion-correction
methods have been used.
The ``PBE-vdW'' results were obtained using the PAW method with the
optB86b vdW-DF \cite{Klime_2009}.
The other vdW-DF data \cite{SI} are similar to the optB86b results
shown.
\label{fig:ingax_2layer}}
\end{figure*}

\begin{figure*}[!htbp]
\centering
\includegraphics[clip,scale=0.58]{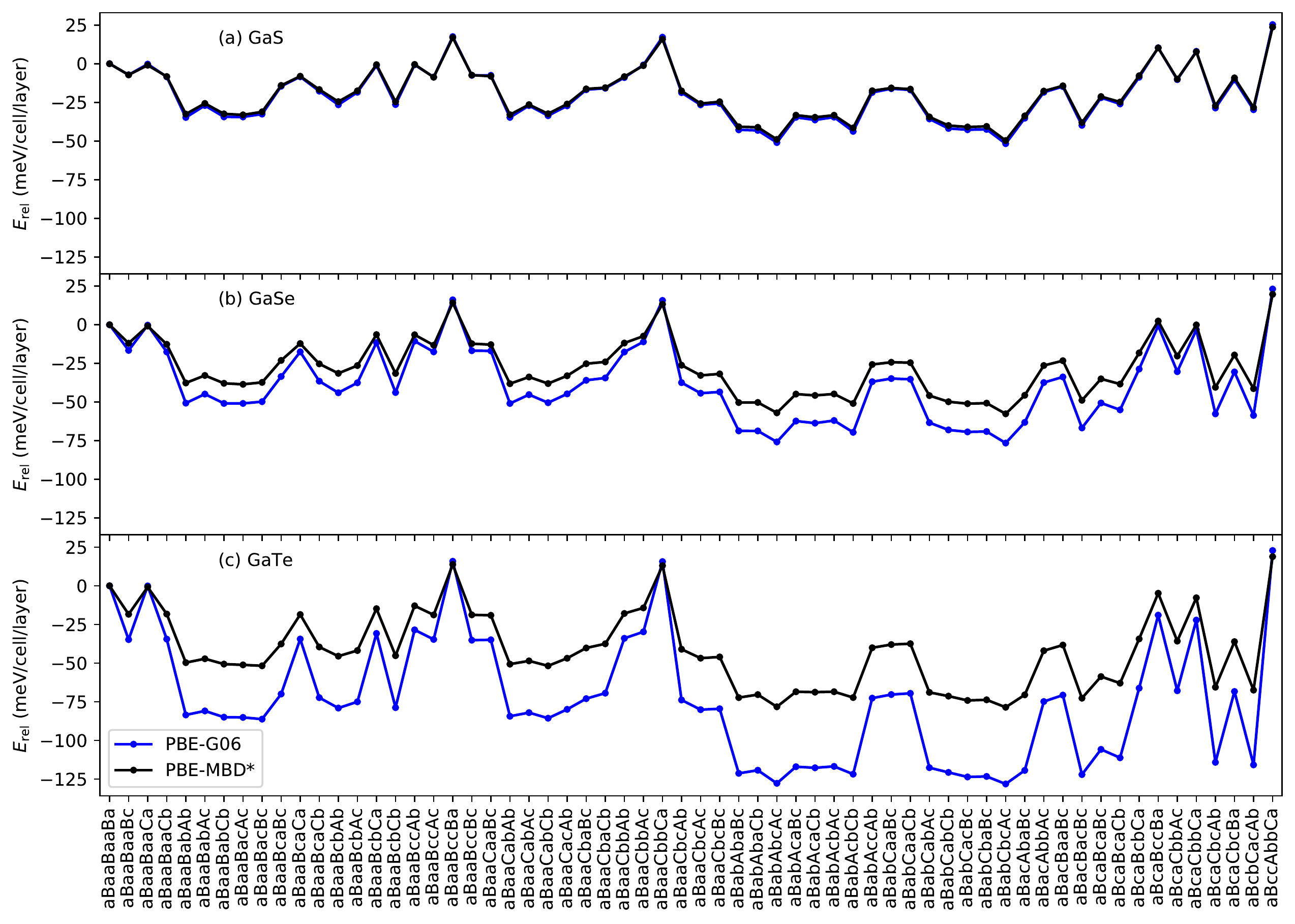}
\includegraphics[clip,scale=0.58]{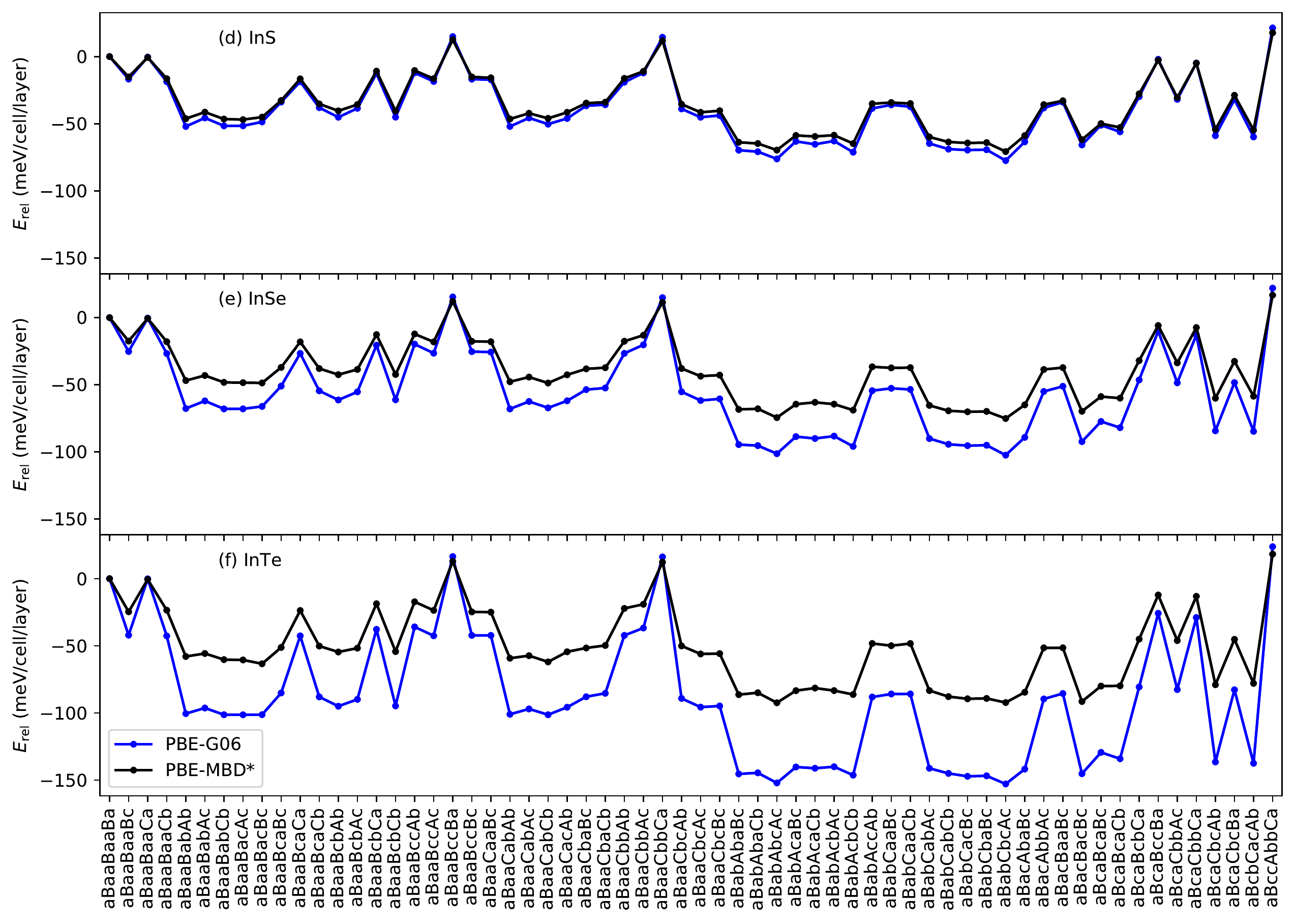}
\caption{DFT energy $E_{\rm rel}$ of three-layer structures relative
  to the AA-stacked $\alpha_{\rm M}$ polytype [\texttt{aBaaBaaBa}
    ($={}$\texttt{aBa})], for (a) GaS, (b) GaSe, (c) GaTe, (d) InS,
  (e) InSe, and (f) InTe.
Different dispersion-correction methods are used.
\label{fig:ingax_3layer}}
\end{figure*}

We obtained a complete set of DFT-PBE-G06 and DFT-PBE-MBD*
total-energy results for all two- and three-layer structures and
fitted Eq.\ (\ref{eq:Etot_fit}) to the data.
We also obtained DFT-LDA, DFT-PBE, DFT-LDA-OBS, and vdW-DF data for
all two-layer structures, to assess the performance of these
functionals.
The DFT results for two-layer structures are shown in
Fig.\ \ref{fig:ingax_2layer}.
The corresponding results for three-layer structures are shown in
Fig.\ \ref{fig:ingax_3layer}.
The disagreements between different dispersion corrections indicate
the limitations of DFT in studies of layered structures.
Alternative methods such as quantum Monte Carlo approaches are
required to provide independent benchmarks \cite{Mostaani_2015}.
We regard the DFT-PBE-MBD* method as somewhat more reliable than the
other dispersion corrections because it describes many-body
interactions and screening effects beyond a description by pairwise
interatomic potentials, and because it has been extensively
benchmarked against diffusion quantum Monte Carlo data
\cite{Ambrosetti_2014b}.

\section{Structural stability \label{sec:stability}}

\subsection{Zero external pressure}

Using Eq.\ (\ref{eq:Etot_fit}) together with the parameters shown in
Table \ref{table:MBDstar_fitting_parameters}, we find that for GaS,
GaSe, GaTe, and InS the most stable hexagonal structure is
\texttt{aBabAb}, which corresponds to the $\beta$ polytype described
in experiments \cite{Kuhn_1976}.
This structure consists of an AA$'$ stacking of $\alpha_{\rm
  M}$-polytype monolayers and has D$_{6h}$ point group and P$6_3$/mmc
space group.
For GaSe, this structure is more stable than the $\varepsilon$ and
$\gamma$ polytypes by $0.86$ meV per monolayer unit cell, and it is
more stable than the $\delta$ polytype by a mere $0.43$ meV per
monolayer unit cell.
The most energetically stable structures of GaSe with unit cells of up
to six layers are shown in Table \ref{table:GaSe_energies}.

\begin{table}[!htbp]
\centering
\caption{Most energetically competitive structures of GaSe with up to
  six layers in the unit cell, together with some other structures of
  interest.
  The DFT-PBE-MBD* energy of each structure relative to \texttt{aBa}
  is shown.
\label{table:GaSe_energies}}
\begin{tabular}{lc}
\hline \hline

Structure & Energy (meV/cell/layer) \\

\hline

 \texttt{aBabAb} ($\beta$-GaSe)  & $-59.416$ \\

 \texttt{aBabAbaBabAbaBabCb}     &  $-59.130$ \\

 \texttt{aBabAbaBabAbaBacAc}     &  $-59.130$ \\

 \texttt{aBabAbaBabAbaCacAc}     &  $-59.130$ \\

 \texttt{aBabAbaBabCbcBcbCb}     &  $-59.130$ \\

 \texttt{aBabAbaCacAc} ($\delta$-GaSe) & $-58.988$ \\

 \texttt{aBabAbaBabCb}           &  $-58.988$ \\

 \texttt{aBabAbaBacAc}           &  $-58.988$ \\

 \texttt{aBabAbaBabCbcAc}        &  $-58.902$ \\

 \texttt{aBabAbaCacAcbCb}        &  $-58.902$ \\

 \texttt{aBabAbaBabCbaBabCb}     &  $-58.845$ \\

 \texttt{aBabAbaBabCbaBacAc}     &  $-58.845$ \\

 \texttt{aBabAbaBabCbcAcbCb}     &  $-58.845$ \\

 \texttt{aBabAbaBacAcaBacAc}     &  $-58.845$ \\

 \texttt{aBabAbaBacAcbCbcAc}     &  $-58.845$ \\

 \texttt{aBabAbaCabAbaBabCb}     &  $-58.845$ \\

 \texttt{aBabAbaCabAbaBacAc}     &  $-58.845$ \\

 \texttt{aBabAbaCabAbaCacAc}     &  $-58.845$ \\

 \texttt{aBabAbaCabAbcBcbCb}     &  $-58.845$ \\

 \texttt{aBabAbaCacAcbCbcAc}     &  $-58.845$ \\

 \texttt{aBabAbaCacBcbCbcAc}     &  $-58.845$ \\

 \texttt{aBabAbcBcbAbaBacAc}     &  $-58.845$ \\

$\vdots$ & $\vdots$ \\

\texttt{aBabCb} ($\varepsilon$-GaSe) & $-58.559$ \\

\texttt{aBabCbcAc} ($\gamma$-GaSe) & $-58.559$ \\

$\vdots$ & $\vdots$ \\

\texttt{aBacBc} & $-56.010$ \\

\hline \hline
\end{tabular}
\end{table}

On the other hand, for InSe and InTe the most stable hexagonal
structure is \texttt{aBacBc}.
This consists of an AB$'$ stacking of $\alpha_{\rm M}$-polytype monolayers.
For InSe this structure is more stable than the $\varepsilon$ and
$\gamma$ polytypes (\texttt{aBabCb} and \texttt{aBabCbcAc}) by just
$0.17$ meV per monolayer unit cell.
The most stable structure differs from that of the gallium
chalcogenides and indium sulfide by a horizontal translation of every
second layer.
Nevertheless, this structure also has D$_{6h}$ point group and
P$6_3$/mmc space group.
The most stable structures of InSe in unit cells of up to six layers
are shown in Table \ref{table:InSe_energies}.

\begin{table}[!htbp]
\centering
\caption{Most energetically competitive structures of InSe with up to
  six layers in the unit cell, together with some other structures of
  interest.
The DFT-PBE-MBD* energy of each structure relative to \texttt{aBa} is
shown.
\label{table:InSe_energies}}
\begin{tabular}{lc}
\hline \hline

Structure & Energy (meV/cell/layer) \\

\hline

 \texttt{aBacBc}  &    $-76.943$ \\

 \texttt{aBabCbaBacBcaBacBc}   &    $-76.888$ \\

 \texttt{aBabCbaCabCbaBacBc}   &    $-76.888$ \\

 \texttt{aBabCbaCabCbaCabCb}   &    $-76.888$ \\

 \texttt{aBabCbaCabCbaCacBc}   &    $-76.888$ \\

 \texttt{aBabCbaBacBc}         &    $-76.860$ \\

 \texttt{aBabCbaCabCb}         &    $-76.860$ \\

 \texttt{aBabCbaCacBc}         &    $-76.860$ \\

 \texttt{aBabCbaCabAbcAc}      &    $-76.844$ \\

 \texttt{aBabCbaCabCbcAc}      &    $-76.844$ \\

 \texttt{aBabCbaBabCbaBacBc}   &    $-76.832$ \\

 \texttt{aBabCbaBabCbaCabCb}   &    $-76.832$ \\

 \texttt{aBabCbaBabCbaCacBc}   &    $-76.832$ \\

 \texttt{aBabCbaBacAcaBacBc}   &    $-76.832$ \\

 \texttt{aBabCbaBacAcbAbcAc}   &    $-76.832$ \\

 \texttt{aBabCbaBacAcbAbcBc}   &    $-76.832$ \\

 \texttt{aBabCbaBacBcaCacBc}   &    $-76.832$ \\

 \texttt{aBabCbaBacBcbAbcBc}   &    $-76.832$ \\

 \texttt{aBabCbaCabAbaCabCb}   &    $-76.832$ \\

 \texttt{aBabCbaCabAbaCacBc}   &    $-76.832$ \\

 \texttt{aBabCbaCabCbaBacAc}   &    $-76.832$ \\

 \texttt{aBabCbaCacBcbAbcAc}   &    $-76.832$ \\

~~~$\vdots$ & $\vdots$ \\

\texttt{aBabCb} ($\varepsilon$-InSe) & $-76.777$ \\

\texttt{aBabCbcAc} ($\gamma$-InSe) & $-76.777$ \\

~~~$\vdots$ & $\vdots$ \\

\texttt{aBabAbaCacAc} ($\delta$-InSe) & $-76.020$ \\

~~~$\vdots$ & $\vdots$ \\

\texttt{aBabAb} ($\beta$-InSe) & $-75.264$ \\

\hline \hline
\end{tabular}
\end{table}

\begin{figure}[!htbp]
\centering
\includegraphics[clip,width=0.4\textwidth]{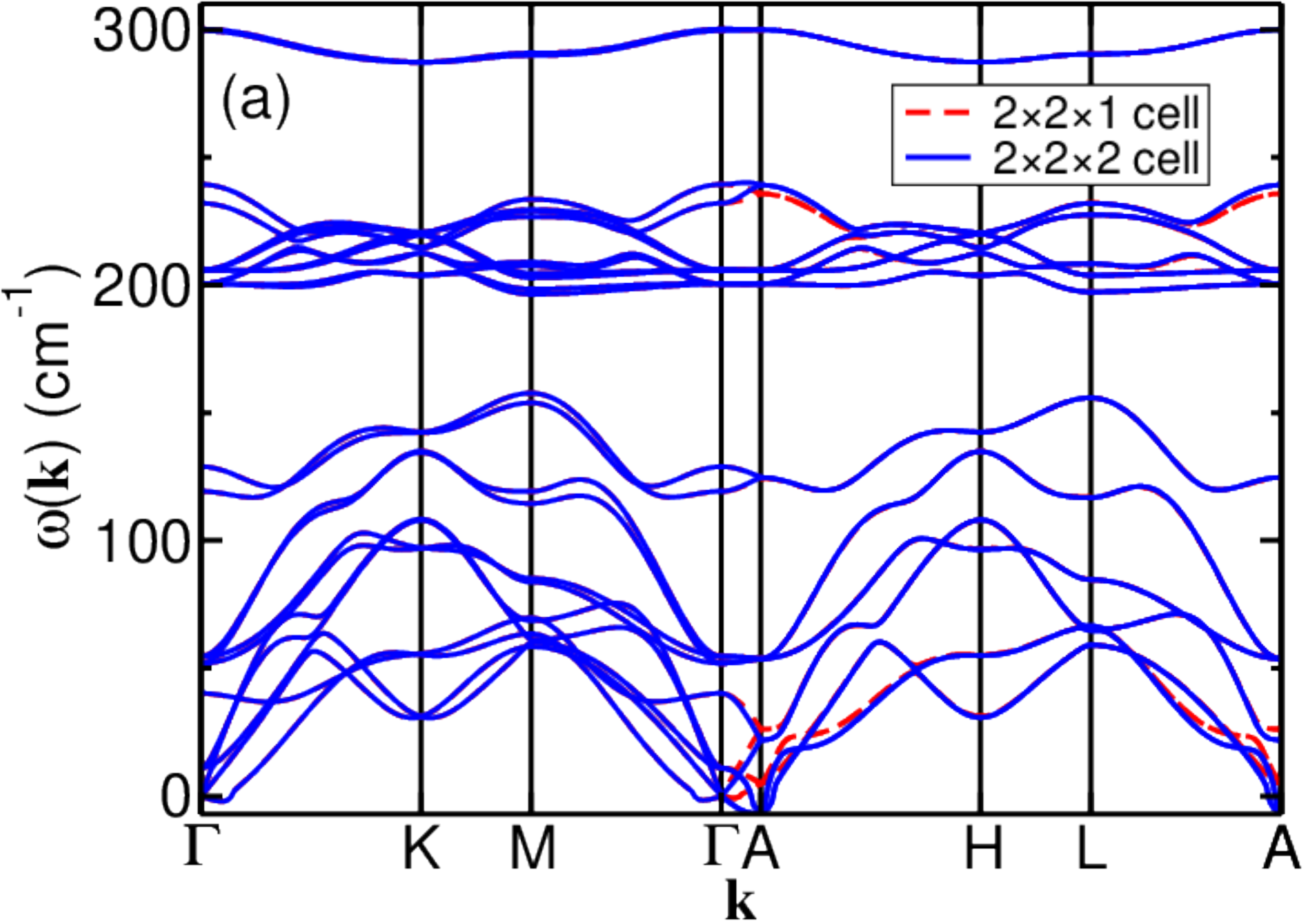}
\\[1ex]
\includegraphics[clip,width=0.4\textwidth]{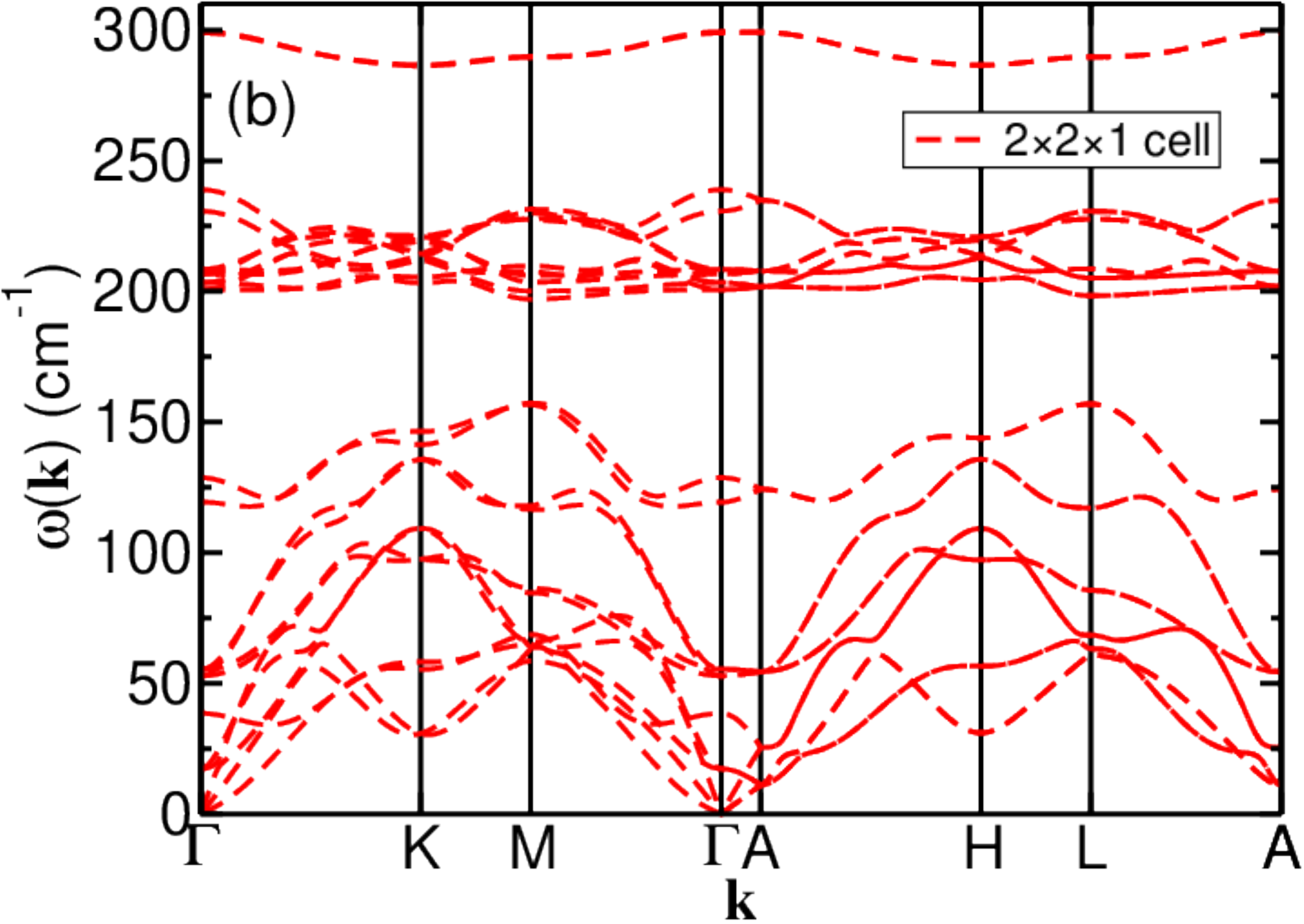}
\caption{DFT-PBE-MBD* phonon dispersion curves of (a) \texttt{aBabAb}
  (the $\beta$ polytype) GaSe and (b) \texttt{aBacBc} GaSe.
The results were obtained using the method of finite displacements in
different sizes of supercell.
 \label{fig:gase_aBabAb_phonons}}
\end{figure}

As a test, we have relaxed the structures of \texttt{aBabAb} (the
$\beta$ polytype) and \texttt{aBabCb} (the $\varepsilon$ polytype)
GaSe without any symmetry constraints.
The initial lattice vectors and atom positions were randomly offset by
a small amount from their exact hexagonal-cell values, and the
positions and lattice vectors were relaxed within DFT-PBE-MBD* at zero
pressure.
This did not lead to a lowering of the total energy relative to the
hexagonal cell, thus providing direct evidence in support of our
assumption that the unit cell is hexagonal in all cases and that the
atoms lie in horizontal sublayers on hexagonal sublattice sites.
Direct confirmation that the structures that we have found to be most
energetically stable in any of the PTMCs are also dynamically stable
is provided by the DFT-PBE-MBD* phonon dispersion curves shown in
Fig.\ \ref{fig:gase_aBabAb_phonons}.
On the other hand, it is known that a monoclinic structure of GaTe is
more stable than the hexagonal structures studied here
\cite{JulienPouzol_1979,Brudnyi_2015}.

While experimental results
\cite{Kuhn_1975,Kuhn1975b,Kuhn_1976,Rigoult_1980,Grimaldi_2020}
support our determination of the $\alpha_{\rm M}$ monolayer structure
as the most stable form, and also agree that the most stable structure
of GaS is \texttt{aBabAb} (the $\beta$ polytype) \cite{Kuhn_1976}, for
InSe and GaSe the \texttt{aBacBc} structure calculated to be most
stable is not one of the commonly observed structures in experiments.
Specifically, the experimental work on InSe finds most often the
$\gamma$ polytype (\texttt{aBabCbcAc}) \cite{Rigoult_1980} and
occasionally the $\varepsilon$ polytype (\texttt{aBabCb})
\cite{Grimaldi_2020}, neither of which has inversion symmetry.
For GaSe the $\varepsilon$ polytype (\texttt{aBabCb}) \cite{Kuhn1975b}
and the $\delta$ polytype (\texttt{aBabCbcBcbAb}) \cite{Kuhn_1975} are
reported, against our result of \texttt{aBabAb} (the $\beta$
polytype).
It should be noted that our results show several structures for each
PTMC of comparable stability on a sub-meV-per-monolayer-unit-cell
scale.
This has important consequences, not only on the theoretical side,
with the structure returned as the most stable being sensitive to the
van der Waals functional chosen, but also on the experimental side,
suggesting that the polytype of a PTMC crystal must be highly sensitive
to the crystal growth conditions.
Indeed, it supports the observation of multiple stacking faults and
regions of different polytypes within a single sample
\cite{DeBlasi_1990}, and suggests that the synthesis of different PTMC
polytypes should be possible with careful tuning of experimental
conditions.
On the theoretical side, an important conclusion is that a
computational method with an accuracy and precision of around $0.1$
meV per monolayer unit cell is required to determine the most stable
PTMC structure reliably.
The $>10$ meV per monolayer unit cell spread of DFT results with
different van der Waals correction schemes and the $\sim 1$ meV per
monolayer unit cell disagreement between independently relaxed
equivalent two- and three-layer structures, together with the
disagreements with experiment regarding the most stable structures,
demonstrate that dispersion-corrected DFT is not currently capable of
such accuracy and precision.

\begin{figure}[!htbp]
\centering
\includegraphics[clip,width=0.4\textwidth]{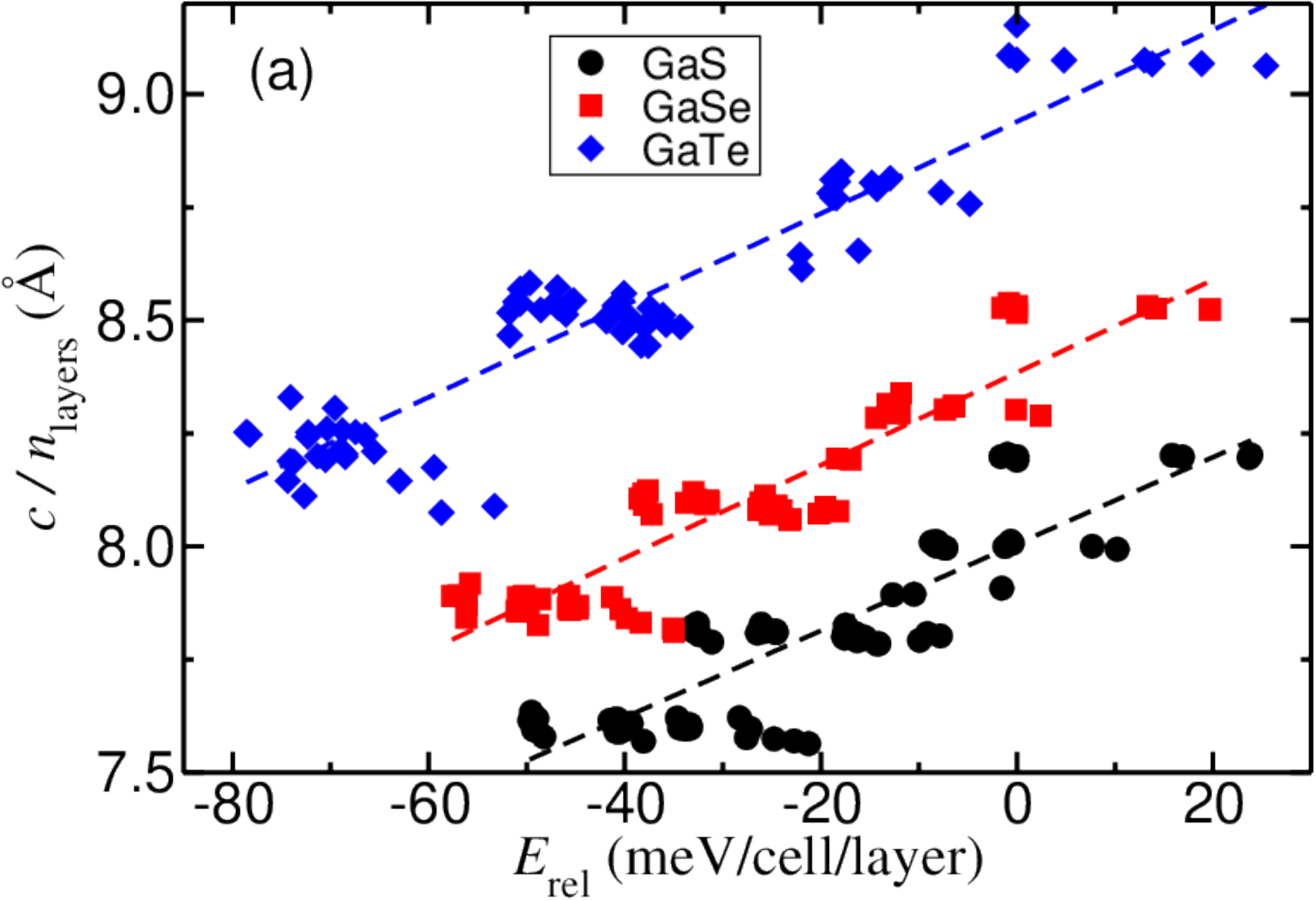} \\[1em]
\includegraphics[clip,width=0.4\textwidth]{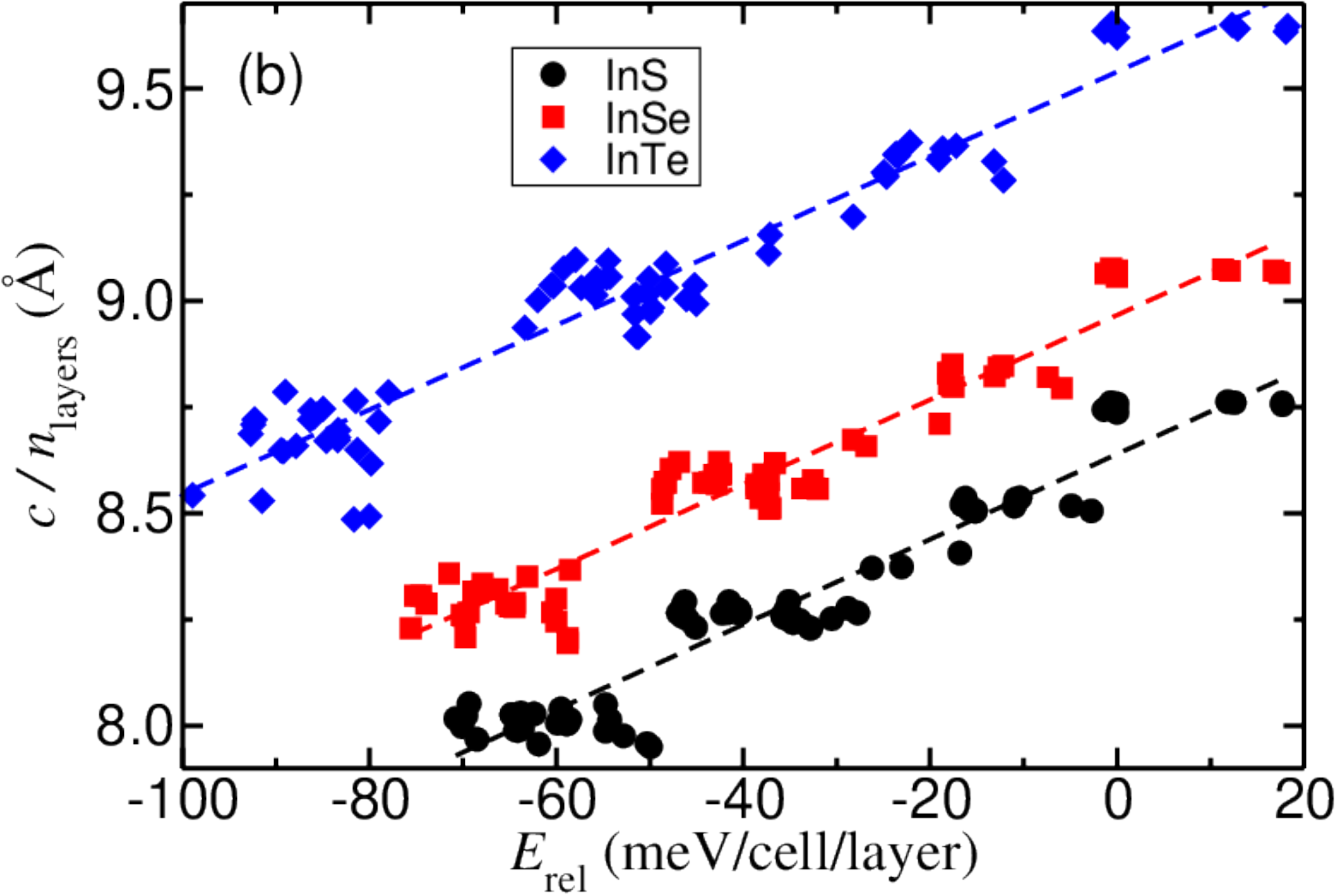}
\caption{Hexagonal lattice parameter $c$ divided by number of layers
  $n_\text{layers}$ against ground-state total energy $E_{\rm rel}$
  for DFT-PBE-MBD*-optimized structures of (a) bulk GaS, GaSe, and
  GaTe and (b) bulk InS, InSe, and InTe.
In each case the ground-state total energy $E_{\rm rel}$ is plotted
relative to that of the \texttt{aBa} structure.
The dashed lines show linear fits to $c/n_\text{layers}$ against
energy for each material.
 \label{fig:sep_v_E}}
\end{figure}

We compare our relaxed lattice parameters with both previous DFT results
and experimental results in Table \ref{table:lattice_params}.
Where comparison is possible, our dispersion-corrected DFT-PBE
calculations agree with experimental results to within $0.2$ {\AA}
(often an order of magnitude better).
The hexagonal $a$ lattice parameter is almost the same for all
structures of a given PTMC, reflecting the in-plane rigidity of the
individual layers.
However, the $c$ lattice parameter is much more sensitive to the
structure, as shown in Fig.\ \ref{fig:sep_v_E}.
High-energy structures generally have larger lattice parameters $c$.

\begin{table*}[!htbp]
\centering
\caption{Hexagonal lattice parameters $a$ and $c$ of $\beta$-GaS
  (\texttt{aBabAb}), $\varepsilon$-GaSe (\texttt{aBabCb}), and
  $\gamma$-InSe (\texttt{aBabCbcAc}) obtained using various methods.
Results without citation were obtained in the present work.
\label{table:lattice_params}}
\begin{tabular}{lcccccc}
\hline \hline

& \multicolumn{2}{c}{$\beta$-GaS (\texttt{aBabAb})} &
\multicolumn{2}{c}{$\varepsilon$-GaSe (\texttt{aBabCb})} &
\multicolumn{2}{c}{$\gamma$-InSe (\texttt{aBabCbcAc})} \\

\raisebox{1.5ex}[0pt]{Method} & $a$ ({\AA}) & $c$ ({\AA}) & $a$ ({\AA}) & $c$ ({\AA}) & $a$ ({\AA}) & $c$ ({\AA}) \\

\hline

DFT-LDA & 3.541 & 15.214 & 3.762 & 15.666 & & \\

DFT-LDA-OBS & 3.517 & 14.939 & 3.695 & 15.346 & & \\

DFT-PBE & 3.633 \cite{Brudnyi_2015}, 3.626 & 16.677
\cite{Brudnyi_2015}, 17.633 & 3.823 \cite{Brudnyi_2015}, 3.811 &
17.848 \cite{Brudnyi_2015}, 18.201 & 4.091 \cite{Brudnyi_2015} &
26.982 \cite{Brudnyi_2015} \\

DFT-PBE-G06 & 3.570 & 15.497  & 3.740 & 15.899 & 3.942 & 25.251 \\

DFT-PBE-MBD* & 3.583 & 15.266 & 3.771 & 15.744 & 4.031 & 24.919 \\

vdW-DF2-C09 & 3.575 \cite{Brudnyi_2015} & 15.460 \cite{Brudnyi_2015} &
3.761 \cite{Brudnyi_2015} & 15.943 \cite{Brudnyi_2015} & 4.028
\cite{Brudnyi_2015} & 24.996 \cite{Brudnyi_2015} \\

Experiment & 3.587 \cite{Kuhn_1976} & 15.492 \cite{Kuhn_1976} & 3.743
\cite{Cenzual_1991} & 15.919 \cite{Cenzual_1991} & 4.002
\cite{Rigoult_1980} & 24.946 \cite{Rigoult_1980} \\

\hline \hline
\end{tabular}
\end{table*}

\subsection{Nonzero pressure}

\begin{figure*}[!htbp]
\centering
\includegraphics[clip,scale=0.3]{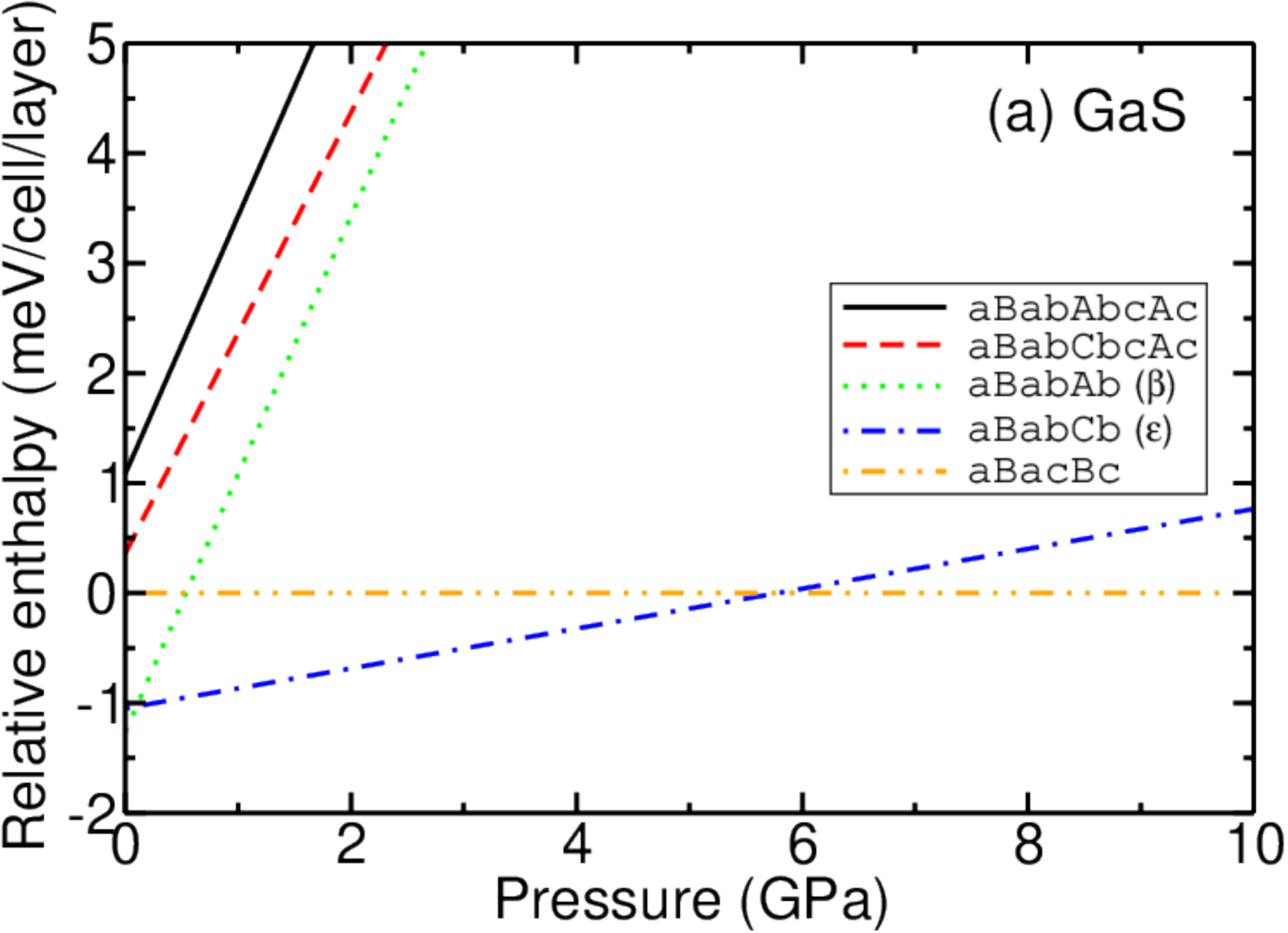} ~~~~~ \includegraphics[clip,scale=0.3]{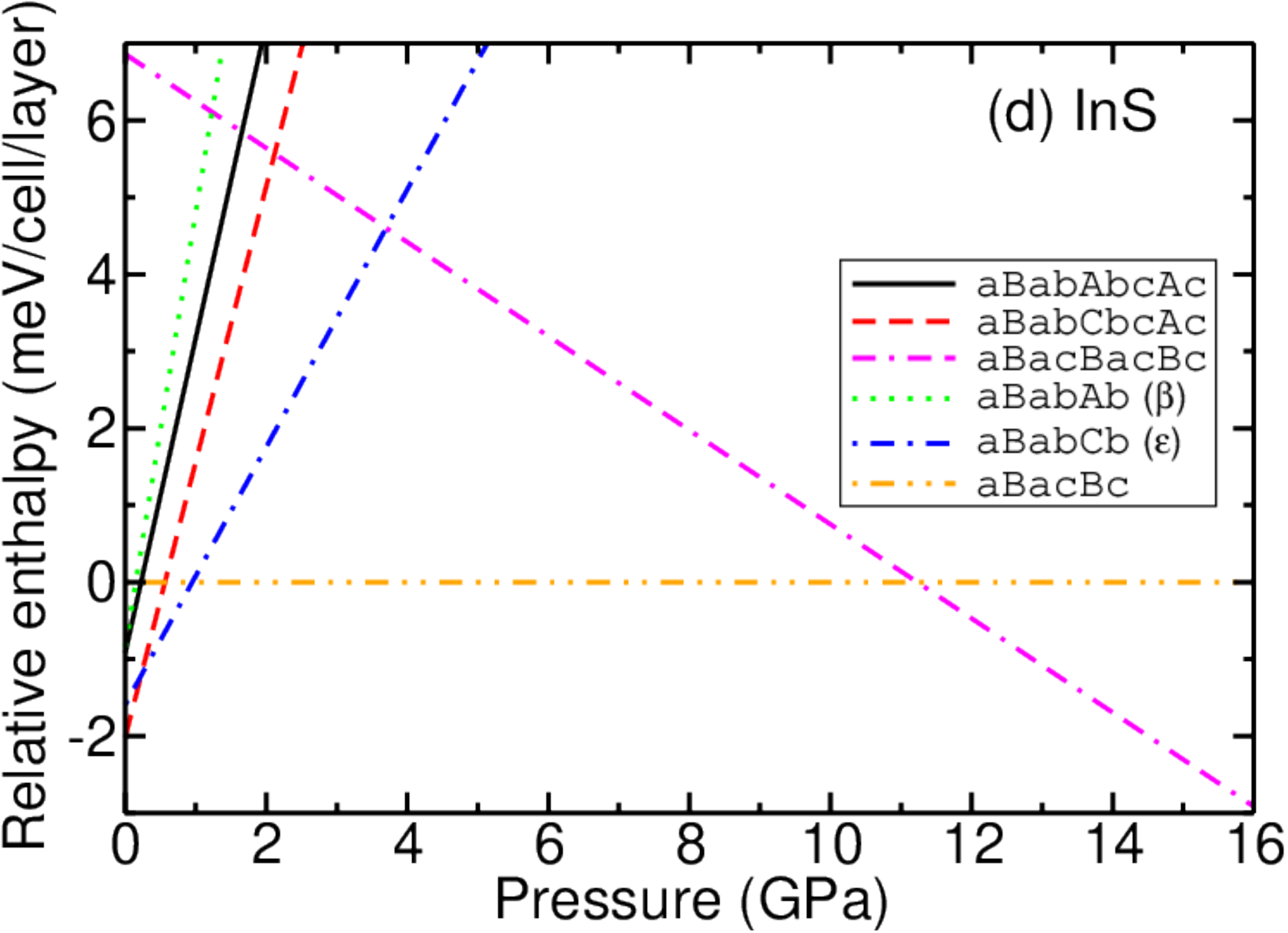} \\[1ex]
\includegraphics[clip,scale=0.3]{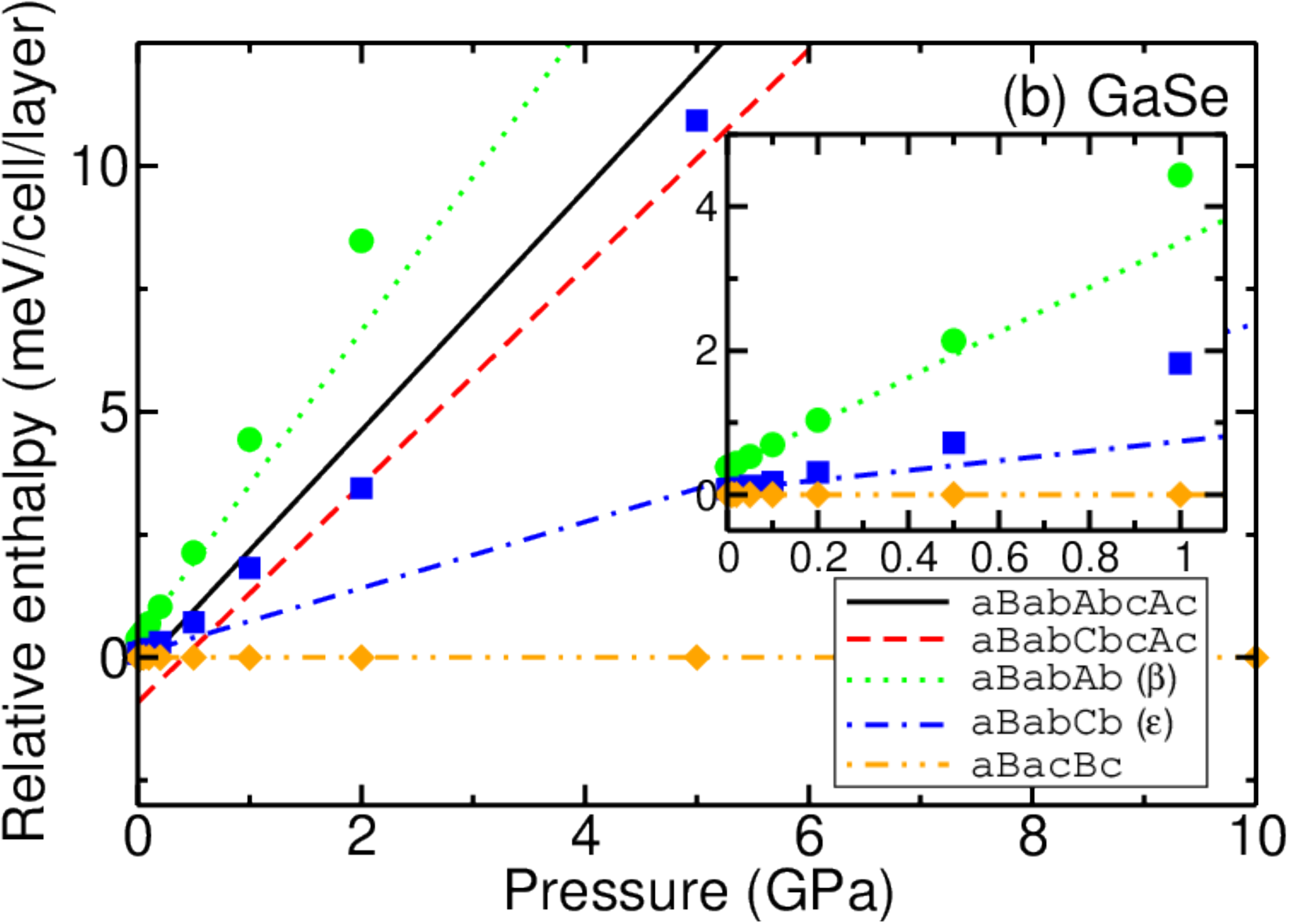} ~~~~~ \includegraphics[clip,scale=0.3]{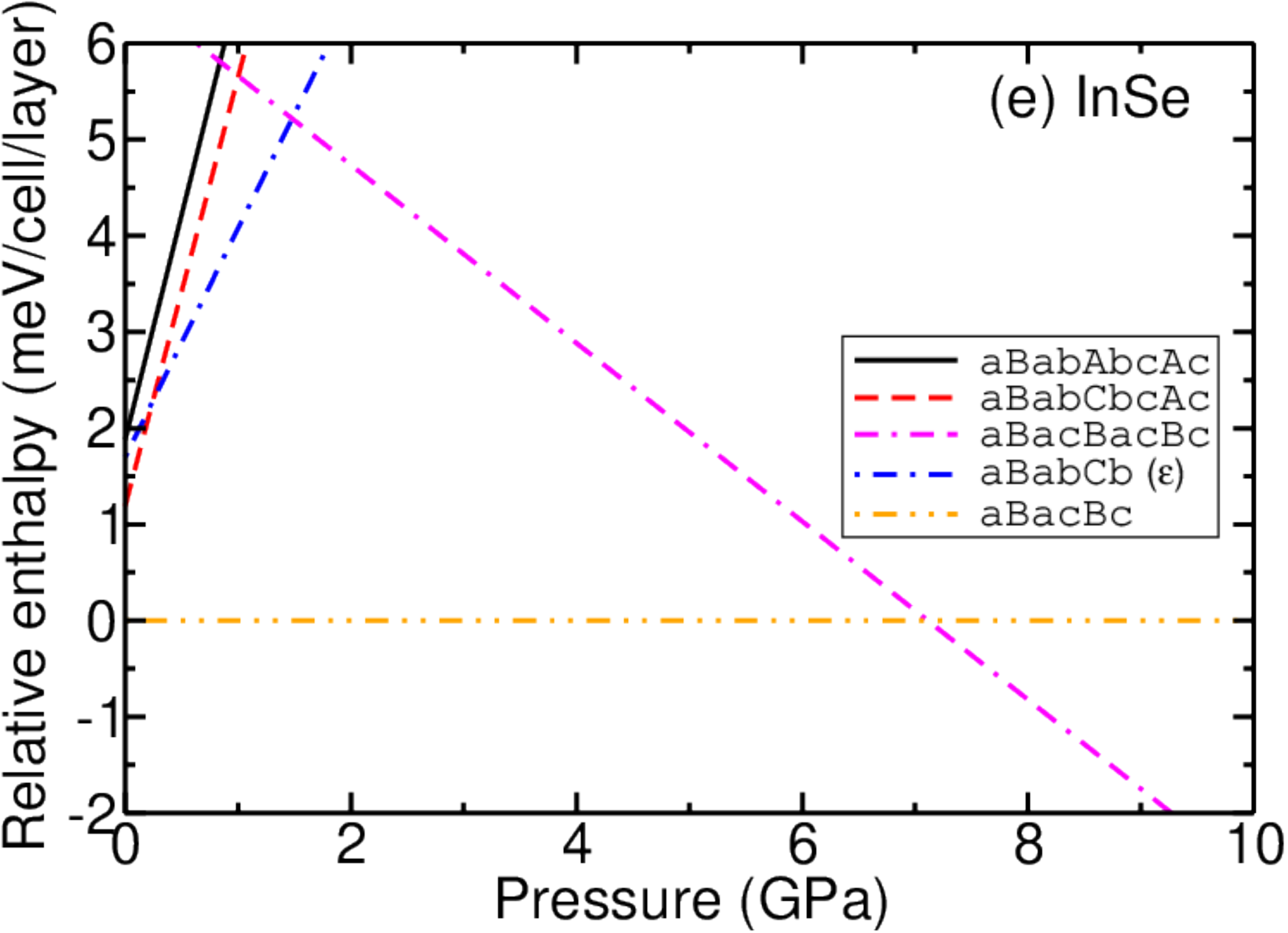} \\[1ex]
\includegraphics[clip,scale=0.3]{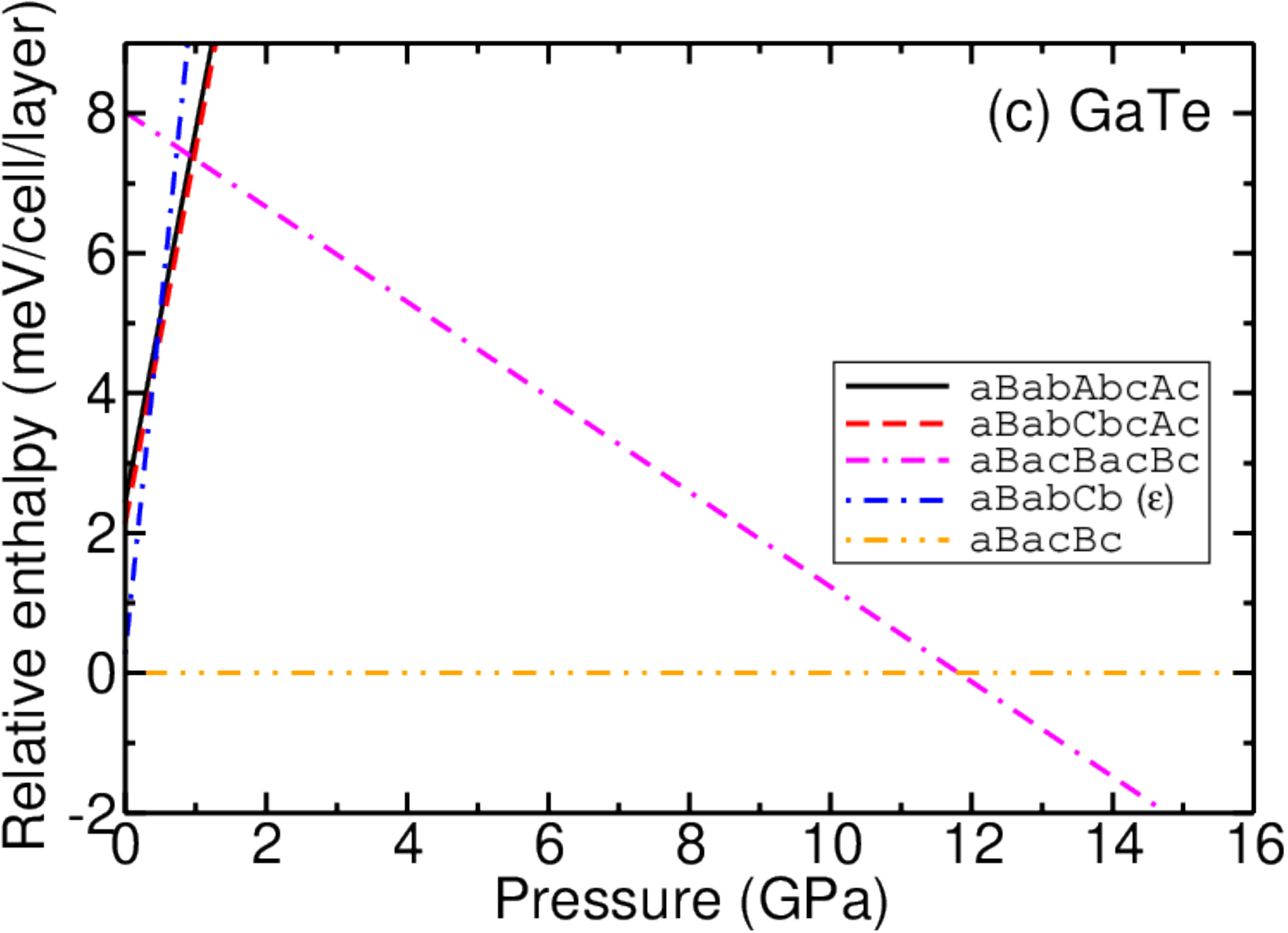} ~~~~~ \includegraphics[clip,scale=0.3]{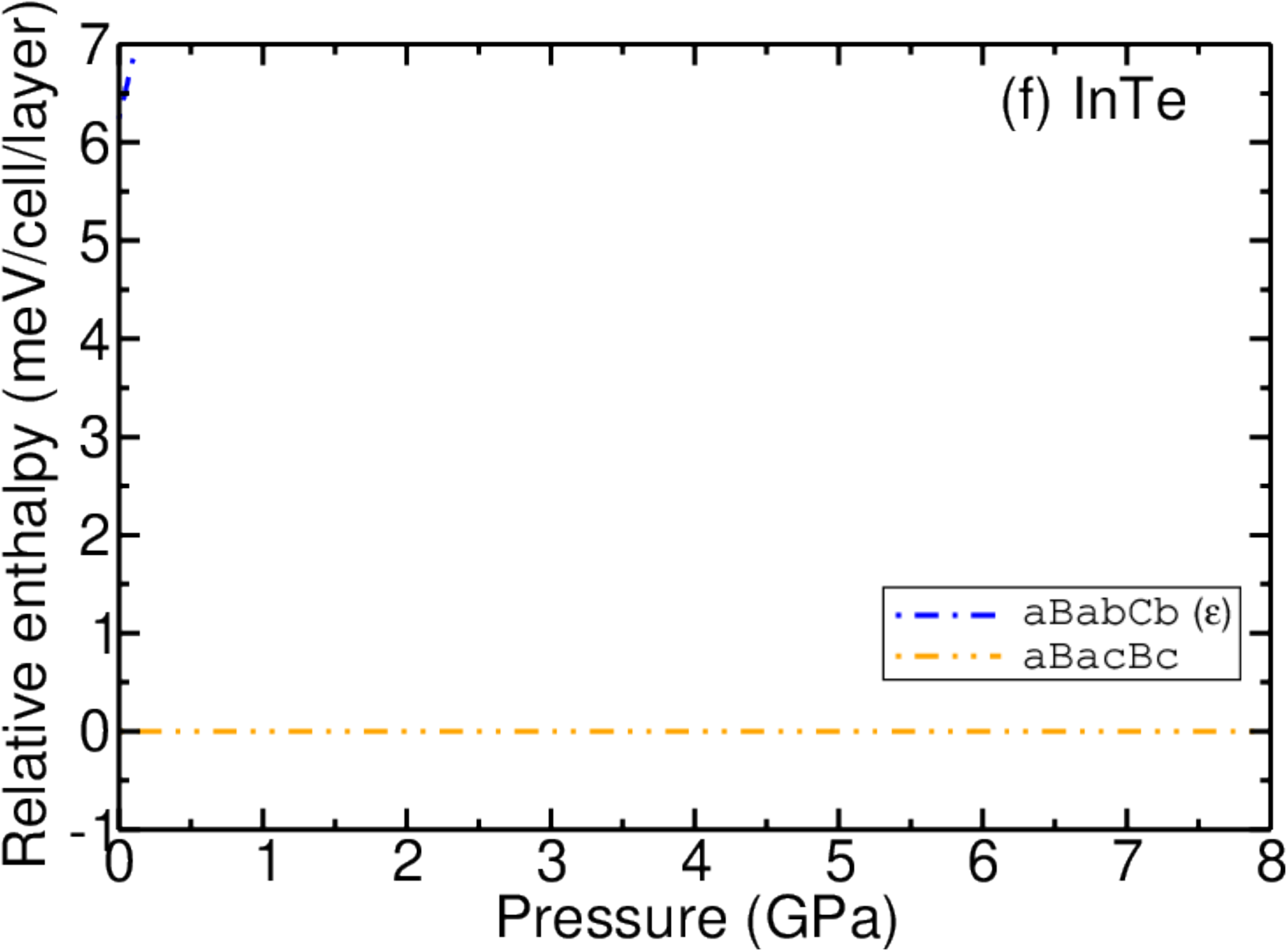}
\caption{Enthalpy against pressure [using Eq.\ (\ref{eq:enthalpy})]
  for energetically competitive structures of (a) GaS, (b) GaSe, (c)
  GaTe, (d) InS, (e) InSe, and (f) InTe.
The zero-pressure energy $E_0$ and volume $V_0$ data were obtained
from DFT-PBE-MBD* calculations.
The enthalpies are plotted relative to the enthalpy of the
\texttt{aBacBc} structure (not the \texttt{aBa} structure).
At any given pressure, the structure with the lowest enthalpy is
thermodynamically favored at zero temperature.
In panel (b) we also show DFT-PBE-MBD* enthalpies obtained directly by
relaxing the structure and lattice parameters at fixed external
pressure.
The inset shows the low-pressure region in greater detail.
Note that the zero-pressure results shown here are obtained directly
from the DFT calculations and do not make use of the fit of
Eq.\ (\ref{eq:Etot_fit}); thus the relative enthalpies shown here
differ from the results shown in Tables \ref{table:GaSe_energies} and
\ref{table:InSe_energies} by around a meV per monolayer unit cell.
\label{fig:H_v_p}}
\end{figure*}

At zero temperature the most thermodynamically stable polytype is the
structure with the lowest enthalpy $H$.
At sufficiently low pressures $p$ we may approximate the enthalpy of a
PTMC structure as
\begin{equation} H \approx E_0+pV_0 + O(p^2), \label{eq:enthalpy} \end{equation}
where $E_0$ and $V_0$ are the zero-pressure energy and volume.
The enthalpy is plotted against pressure for energetically competitive
structures of gallium chalcogenides and indium chalcogenides in
Fig.\ \ref{fig:H_v_p}.
Of the two-layer structures, \texttt{aBabCb} (the $\varepsilon$
polytype), \texttt{aBacBc}, and (in GaS, GaSe, and InS)
\texttt{aBabAb} (the $\beta$ polytype), have DFT-PBE-MBD* energies
within a few meV per monolayer cell of each other at zero pressure,
but at higher pressure, \texttt{aBabAb} (the $\beta$ polytype) is
clearly disfavored.
More generally, the application of pressure simplifies the picture by
reducing the number of competing structures and increasing the
relative enthalpies of those structures.
In GaSe, the three-layer structures \texttt{aBabCbcAc} (the $\gamma$
polytype) and \texttt{aBabAbcAc} are competitive at low pressure.
This is not the case in InSe.
However, at very high pressures, three-layer structures may be
favored in InSe.
Below 7.1 GPa, the \texttt{aBacBc} structure of InSe is favored; above
7.1 GPa, the \texttt{aBacBacBc} structure of InSe is favored.
The structures favored at high pressures feature PTM dimers on the same
sublattice and neighboring chalcogen atoms on different sublattices,
as would be expected from steric considerations.
At low pressure it is once again clear that accuracy and precision of
around $0.1$ meV per monolayer unit cell are required to identify the
most stable polytype unambiguously.

In Fig.\ \ref{fig:H_v_p}(b) we compare the linear approximation to
the enthalpy [Eq.\ \ref{eq:enthalpy})] with DFT enthalpies obtained by
directly relaxing the lattice vectors at a given external pressure.
We find that the linear approximation is of quantitative accuracy on a
meV-per-monolayer-unit-cell scale for relative enthalpies up to around
$1$ GPa.
Beyond this, the linear approximation provides a qualitative picture
that generally preserves the ordering of the structures, at least up to
$\sim 10$ GPa.

In all bulk PTMCs at multi-GPa pressures, the \texttt{aBacBc}
structure is found to be the most stable structure over a broad range
of pressures in DFT-PBE-MBD* calculations.
This is the inversion-symmetric structure that is predicted to be most
stable for InSe and InTe at zero pressure, and consists of
AB$'$-stacked $\alpha_{\rm M}$ monolayers.
Despite its ubiquitous presence in the theoretical calculations, we
are not aware that this structure has previously been reported.

\section{Electronic band structure \label{sec:band_structure}}

\begin{figure}[!htbp]
\centering
\includegraphics[clip,width=0.4\textwidth]{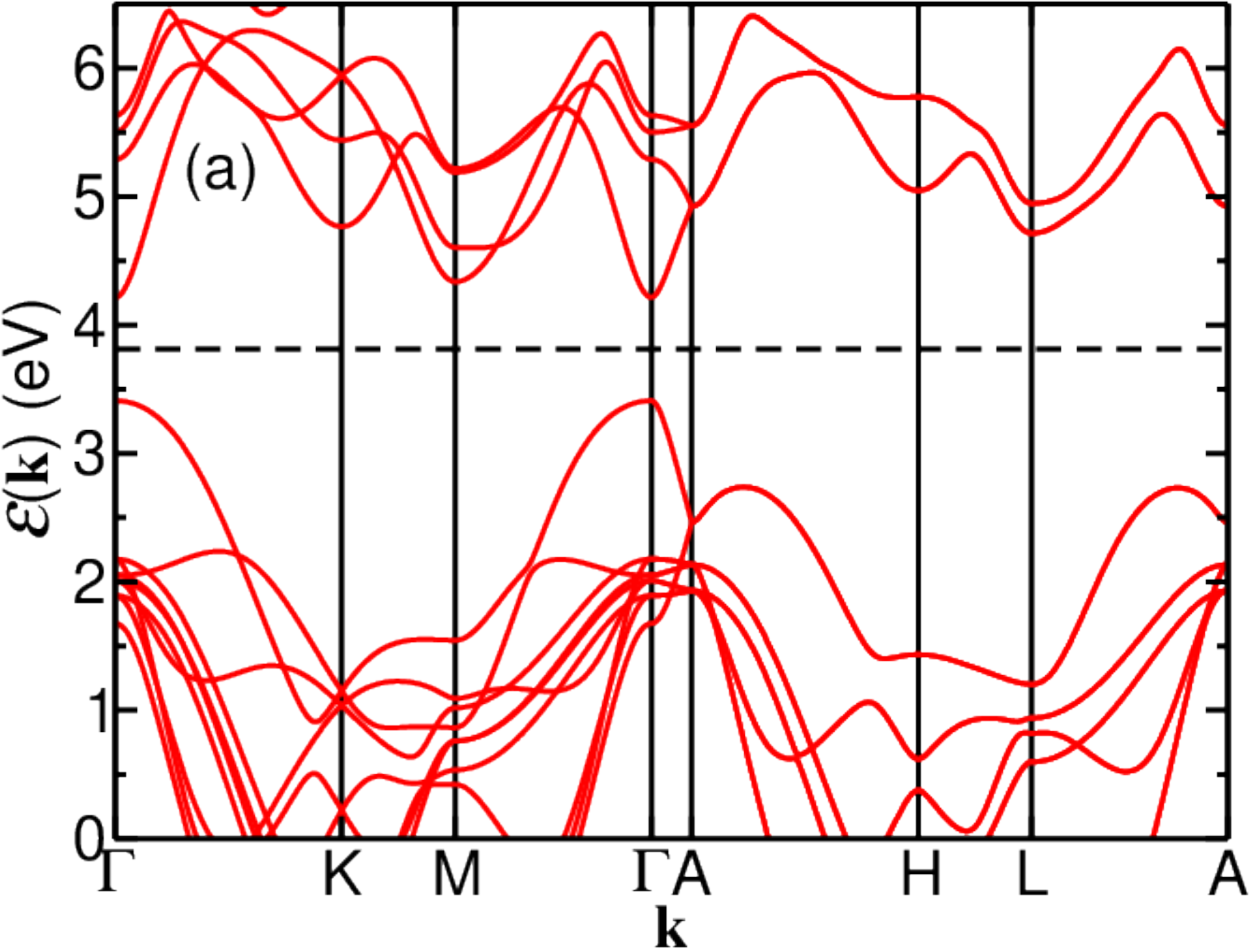}
\llap{\raisebox{2.5cm}{\includegraphics[clip,width=1.75cm]{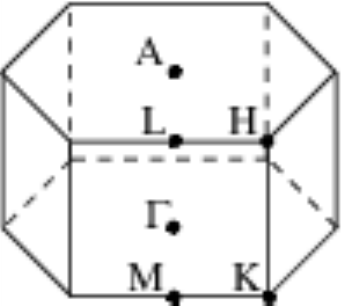}\hspace{0.9cm}}}
\\ \includegraphics[clip,width=0.4\textwidth]{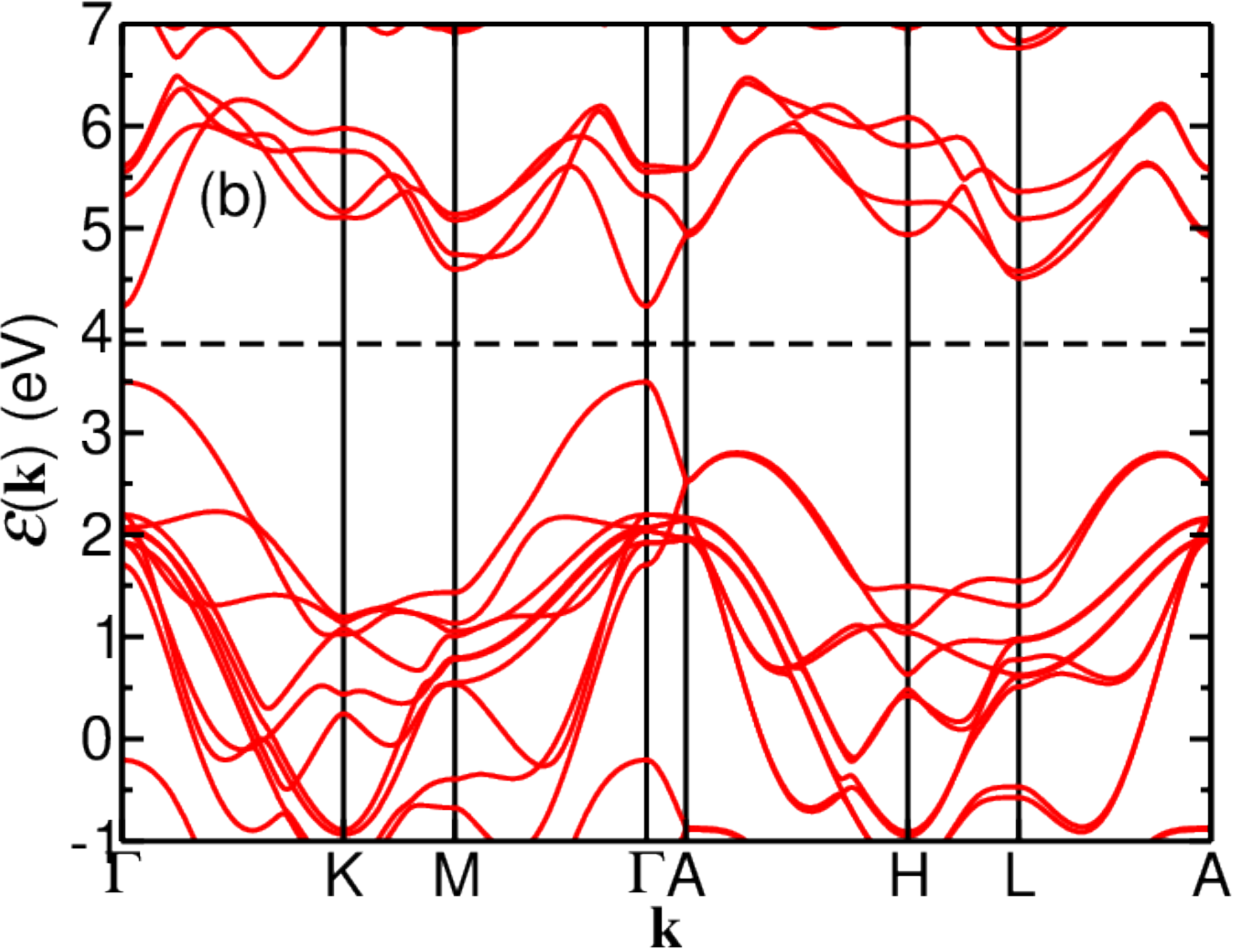}
\\ \includegraphics[clip,width=0.4\textwidth]{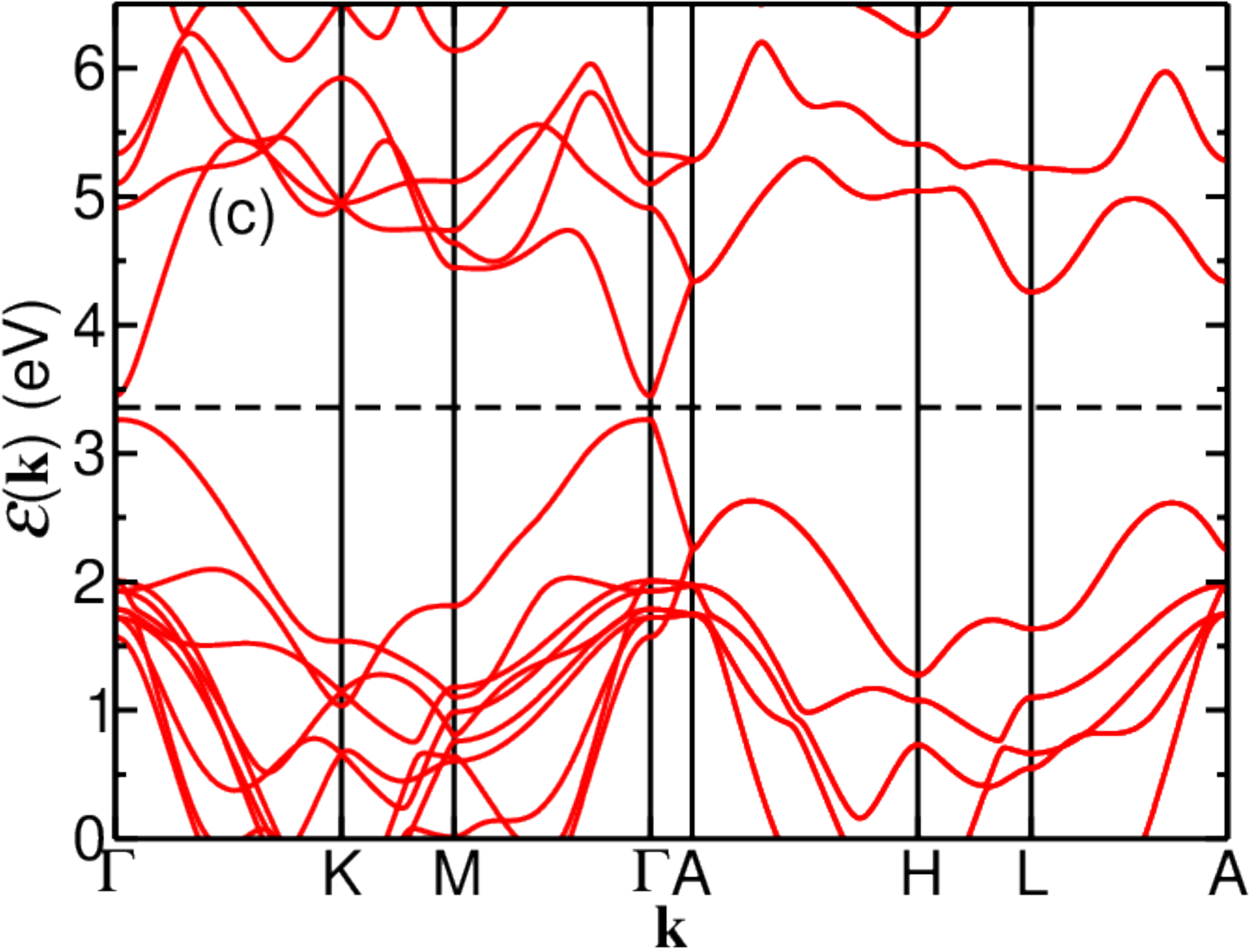}
\caption{Electronic band structures of low-energy structures of bulk
  PTMCs: (a) \texttt{aBabAb} GaSe (the $\beta$ polytype and the
  lowest-energy structure in theory), (b) \texttt{aBabCb} GaSe (the
  $\varepsilon$ polytype), and (c) \texttt{aBacBc} InSe (the
  lowest-energy structure in theory).
The horizontal dashed line shows the Fermi energy in each case.
The inset to panel (a) shows the hexagonal Brillouin zone.
 \label{fig:band_structures}}
\end{figure}

In Fig.\ \ref{fig:band_structures} we plot the DFT-PBE electronic band
structures of the theoretically most stable polytypes of GaSe and InSe
(\texttt{aBabAb} and \texttt{aBacBc}, respectively) and the
experimentally observed \cite{Kuhn1975b} $\varepsilon$ polytype
(\texttt{aBabCb}) of GaSe.
The structures were relaxed using DFT-PBE-MBD*.
In each case the polytypes exhibit a direct band gap at the $\Gamma$
point of the two-layer hexagonal Brillouin zone.
The DFT-PBE band gaps, which are expected to be significant
underestimates of the true gaps \cite{Perdew_1985}, are $0.804$ and
$0.742$ eV for the \texttt{aBabAb} and \texttt{aBabCb} structures of
GaSe, respectively.
The low-energy band structure is qualitatively similar for these two
energetically competitive structures of GaSe.
The DFT-PBE band gap of the most stable structure of InSe
(\texttt{aBacBc}) is much smaller than the gap of GaSe, at $0.183$
eV\@.

\begin{figure}[!htbp]
\centering
\includegraphics[clip,width=0.4\textwidth]{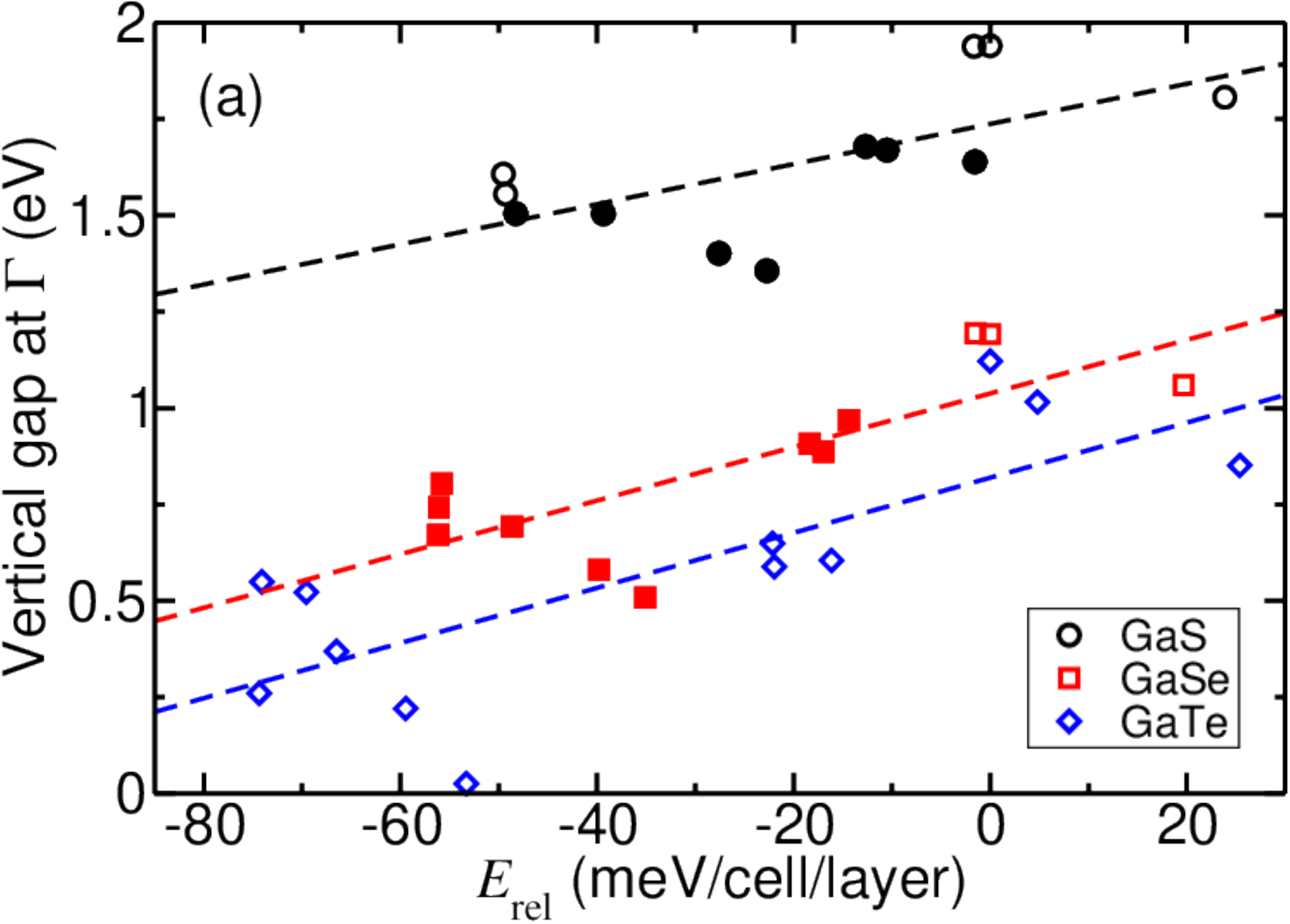} \\[1em]
\includegraphics[clip,width=0.4\textwidth]{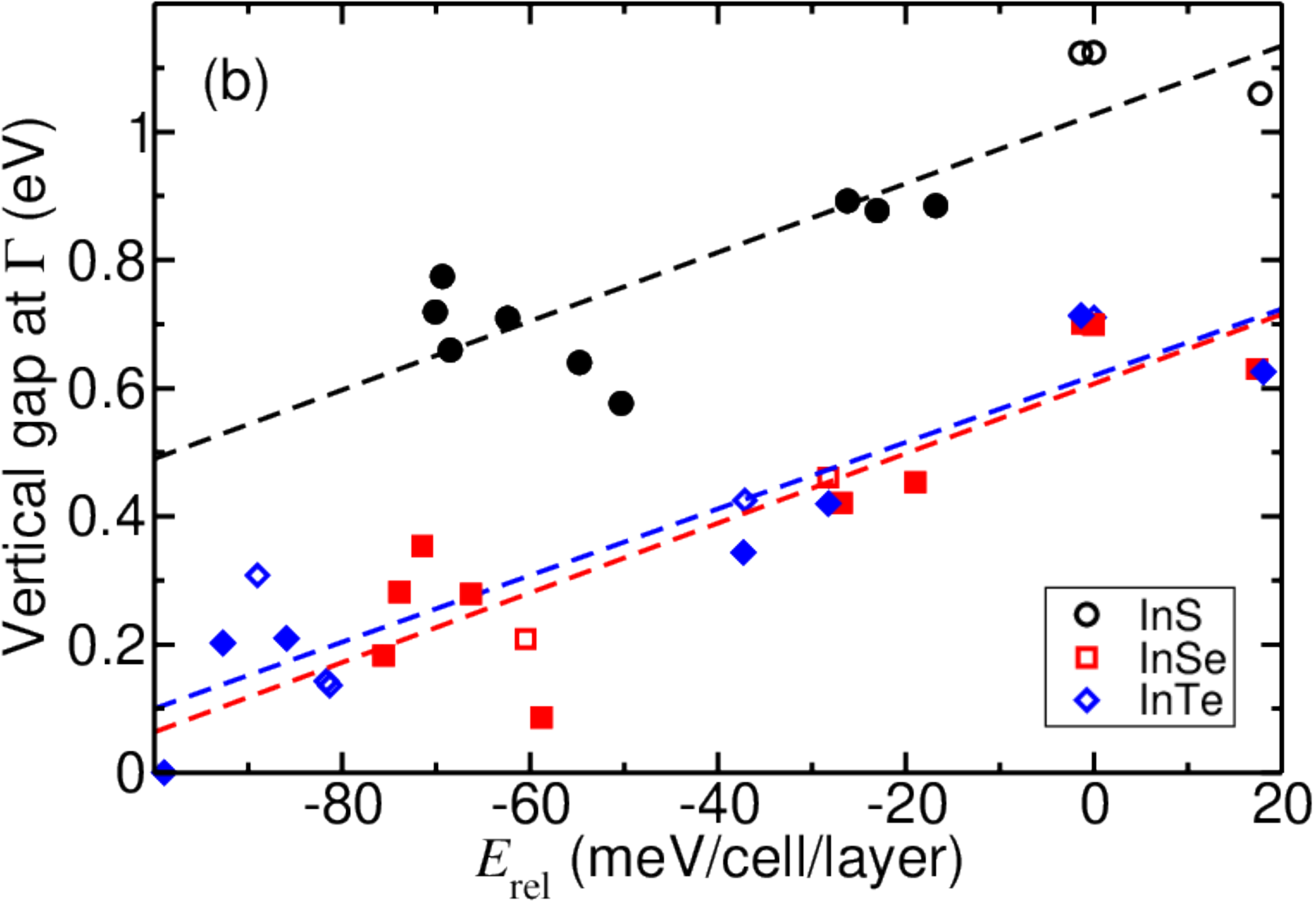}
\caption{Vertical band gap at $\Gamma$ against ground-state total
  energy $E_{\rm rel}$ relative to that of the \texttt{aBa} structure
  for DFT-PBE-MBD*-optimized two-layer structures of (a) bulk GaS,
  GaSe, and GaTe and (b) bulk InS, InSe, and InTe.
The band gaps were calculated using DFT-PBE and the total energies were
evaluated using DFT-PBE-MBD*.
The dashed lines show linear fits to the gap against energy for each
of the three materials.
Where the symbols are filled, the vertical gap at $\Gamma$ is equal to
the fundamental band gap.
 \label{fig:InX_gap_v_E}}
\end{figure}

We have examined the band gap of a range of two-layer structures for
each material, finding that the vertical band gap at $\Gamma$ and the
ground-state energy of each structure are positively correlated,
although with significant noise: see Fig.\ \ref{fig:InX_gap_v_E}.
PTMC structures with smaller band gaps tend to be more stable.
In fact, the most stable two-layer structure of InTe has a direct gap
at $\Gamma$ of just $0.2$ meV\@.
In most cases the vertical gap at $\Gamma$ is the fundamental gap,
especially for low-energy structures.
A notable exception is GaTe, where the vertical gap at $\Gamma$ is
nonfundamental for all the two-layer structures.
In the most stable two-layer GaTe structure, the valence-band maximum
is at $\Gamma$ but the conduction-band minimum is on the $\Gamma$--$M$
line.
Previous work using DFT and many-body perturbation theory has shown
that $\gamma$-InSe changes from a direct-gap material to an
indirect-gap material under high pressure \cite{Ferlat_2002}.

The experimentally measured gaps of $\beta$-InSe, $\gamma$-InSe and
$\epsilon$-InSe are $1.28$ eV \cite{Gurbulak_2014}, $1.25$--$1.29$ eV
\cite{Manjon_2001,Julien_2003}, and $1.4$ eV \cite{Lei_2014},
respectively, which are (as expected) very much larger than the
DFT-PBE InSe gaps of energetically stable polytypes shown in
Fig.\ \ref{fig:InX_gap_v_E}.
Nevertheless, we would expect the qualitative conclusion that the
band gap of a PTMC polytype is positively correlated with its energy to
continue to hold.

We note that the dispersion of the band-edge states in the
out-of-plane direction along $\Gamma$-A is substantial.
The electronic structure is very much three-dimensional, despite the
layered crystalline structure of the PTMCs.
This dispersion arises due to strong interlayer hybridization of $p_z$
orbital states on chalcogen atoms in the band-edge wave
functions \cite{Magorrian_2016}.
It is the restriction of out-of-plane momentum in ultrathin PTMC films
that gives rise to their strong thickness-dependent electronic and
optical properties, with an increase in band gap for a reduced number
of layers \cite{Bandurin_2016,Aziza_2017}.

\section{Conclusions \label{sec:conclusions}}

We have used dispersion-corrected DFT methods to examine the relative
stability of a large number of candidate bulk hexagonal PTMC
polytypes.
For all PTMCs there is a clear consensus among DFT functionals that
the $\alpha_{\rm M}$ monolayer polytype, in which the chalcogen atoms
lie on the same hexagonal sublattice, is $E_{\rm ab}=16$--$24$ meV per
monolayer unit cell more stable than the $\beta_{\rm M}$ polytype, in
which the chalcogen atoms lie on different hexagonal sublattices;
indeed, all experimentally observed bulk polytypes only feature
$\alpha_{\rm M}$ monolayers
\cite{Kuhn_1975,Kuhn1975b,Kuhn_1976,Rigoult_1980,Grimaldi_2020}.
Our DFT-PBE-MBD* calculations show that there is an energy gain of
$-E_{\rm nc}=50$--$100$ meV per monolayer unit cell from having
neighboring chalcogen atoms on different hexagonal sublattices; again,
all experimentally observed polytypes only have neighboring chalcogen
atoms on different hexagonal sublattices
\cite{Kuhn_1975,Kuhn1975b,Kuhn_1976,Rigoult_1980,Grimaldi_2020}.
The DFT-PBE-MBD* energy gain associated with PTM dimers in neighboring
layers lying on different hexagonal sublattices is $-E_{\rm np} =
0.04$--$2.7$ meV per monolayer unit cell.
This leads to a tendency to avoid AA-stacked structures at zero
pressure.
However, in InSe and InTe this is offset by an energy penalty of
$E_{\rm snn}=1.5$--$2.9$ meV per monolayer unit cell associated with
PTM dimers and next-nearest chalcogen atoms lying the same hexagonal
sublattices; it is geometrically impossible to have an AB- or
ABC-stacked $\alpha_{\rm M}$ structure in which PTM dimers and
next-nearest chalcogen atoms all lie on different hexagonal
sublattices.
The interplay between these effects leads to a subtle, sub-meV
competition between polytypes.
Disagreements between dispersion-corrected DFT total energies are of
order $10$ meV per monolayer unit cell.
Disagreements between the relative energies of the lowest-energy
polytypes are of order $1$--$5$ meV per monolayer unit cell.
Only for GaS is the observed stable polytype ($\beta$) predicted by
DFT-PBE-MBD* to have the lowest energy; however, in GaSe and InSe the
observed polytypes are very close in energy to the theoretically most
stable structure.
We conclude that dispersion-corrected DFT methods are not yet able to
predict the relative stability of bulk PTMC polymorphs reliably;
however, they can provide insights into the energy scales involved and
the types of structures that are favored.
The small energy differences between competing polytypes imply that a
wide variety of different polytypes are likely to be found in
experiments, and that stacking faults must be common in PTMC samples.

We find that application of pressure tends to favor an \texttt{aBacBc}
PTMC structure that has not previously been reported.
In fact this polytype is found to be most stable within DFT-PBE-MBD*
at zero pressure for InSe and InTe.
We also find that there is a positive correlation between the
ground-state total energy and the electronic band gap; energetically
stable PTMC polytypes tend to have smaller band gaps.

\begin{acknowledgments}
We acknowledge useful conversations with V.\ I.\ Fal'ko.
All relevant data present in this publication can be accessed at
\textcolor{red}{URL???}.
Computational resources were provided by Lancaster University's
high-end computing facility,
and by the University of Manchester Computational Shared Facility.
SJM acknowledges support from EC Quantum Technology Flagship Project
No.\ 2D-SIPC\@.
\end{acknowledgments}

\appendix

\section{Methodology \label{app:methodology}}

We used DFT as implemented in the \textsc{castep} \cite{Clark_2005}
plane-wave-basis code to compute the relative energies of PTMC crystals
in a variety of bulk hexagonal structures.
We used ultrasoft pseudopotentials to represent atomic cores and we
used plane-wave cutoff energies of at least $566$ eV\@.
The maximum distance between ${\bf k}$ points in the Monkhorst-Pack
grid was less than $0.0189$ {\AA}$^{-1}$ in each case.
The force tolerance for geometry optimization was $0.514$
meV\,{\AA}$^{-1}$.
We verified that near-identical relative energies for PTMC structures
were obtained using the \textsc{vasp} \cite{Kresse_1996} DFT code with
projector augmented-wave (PAW) pseudopotentials instead of
\textsc{castep}.
In the \textsc{vasp} calculations the basis consisted of plane waves
with a cutoff energy of $680$ eV and the Brillouin zone was sampled by
a Monkhorst-Pack grid of $18\times 18\times 4$ points.
The crystals were fully optimized with a force tolerance of $0.005$
eV\,\AA$^{-1}$\@.
We also verified that \textsc{castep} DFT relative energies obtained
using norm-conserving pseudopotentials were in agreement (on a
meV-per-monolayer-unit-cell scale) with our results obtained using
ultrasoft pseudopotentials.
Full data sets can be found in the Supplemental Information \cite{SI}.

\bibliography{bulk_PTMC_DFT}

\end{document}